\documentclass[11pt]{article}

\usepackage{graphicx} 

\usepackage{amsmath}
\usepackage{amsthm} 
\usepackage{amssymb} 
\usepackage{thmtools}

\usepackage{enumerate}
\usepackage{multicol}
\usepackage{multirow}
\usepackage{xcolor}
\usepackage{pifont}
\usepackage{kbordermatrix}
\usepackage{caption}
\usepackage{subcaption}

\newtheorem{prop}{Proposition}
\newtheorem{xmpl}{Example}

\newtheorem{rema}{Remark}

\theoremstyle{definition}
\newtheorem{definition}{Definition}
 
\theoremstyle{remark}

\DeclareMathOperator*{\argmax}{\arg\!\max}

\newcommand{\mmstab}{{\tiny $ \color{orange}{\bigstar} \phantom{ \bigstar \bigstar \bigstar  \bigstar}  $}}

\newcommand{\mstab}{{\tiny $ \color{orange}{\bigstar \bigstar} \phantom{ \bigstar \bigstar \bigstar }  $}}

\newcommand{\nstab}{{\tiny $ \color{orange}{\bigstar \bigstar \bigstar} \phantom{ \bigstar \bigstar }$}}

\newcommand{\pstab}{{\tiny $ \color{orange}{\bigstar  \bigstar  \bigstar  \bigstar } \phantom{ \bigstar}  $}}

\newcommand{\ppstab}{{\tiny $ \color{orange}{\bigstar  \bigstar  \bigstar  \bigstar \bigstar} $}}

\title{Fair Division of Goods in the Shadow of Market Values} 

\author{Marco Dall'Aglio\thanks{Corresponding author. I wish to thank three anonymous referees, whose comments greatly improved the paper. I would also like to thank all the Invited Speakers to the ``De Aequa Divisione'' workshop which took place in Rome, May 23--25, 2019: Anna Bogomolnaia,  Markus Brill, Daniela Di Cagno, Edith Elkind, Vasilis Gkatzelis, J\'er\^ome Lang, Juan D.\ Moreno Ternero,  Herv\'e Moulin, Antonio Nicol\'o, Fedor Sandomirskiy and William Thomson. They all provided feedback during the presentation of of this work at an early stage. Additional thanks go to Herv\`e Moulin for fruitful discussions that took place during the early stage of this work. Finally, I would like to thank Elizabeth Mary Bevan and Cecilia Flori for correcting my language mistakes in the many revisions. All the mistakes as yet undiscovered are mine.}
\\
Luiss University, Italy} 
\date{July 28, 2022}  

\begin{document}

\maketitle 

\begin{abstract}
Inheritances, divorces or liquidations of companies require common assets to be divided among the entitled parties. Legal methods usually consider the market value of goods, while fair division theory takes into account the parties' preferences expressed as utilities. I combine the two practices to define a procedure that optimally allocates divisible goods with market values to people with easily elicited preferences. 
Imposing an exact equality onn the bundles' monetary values may produce unacceptable solutions. I drop the tight requirement and suggest a procedure in which the differences in the monetary  values are explained in terms of satisfaction per monetary share as perceived by the agents. A robustness study shows the consequences of misspecification in the model parameters.

\end{abstract}

\section{Introduction}
Law and mathematics traditionally do not get along very well, but opportunities for interaction are becoming more and more frequent. In September 2017, the author joined the ``Conflict Resolution though Equitable Algorithms'' (CREA) consortium: a two-year project funded by the E-Justice programme of the European Union, which saw the collaboration of law, mathematics and computer science researchers, together with Stakeholder Associations, from eight countries in the European Union. One of the project's purposes was, in the words of the proposal, to use fair division theory ``to introduce new mechanisms of dispute resolution as a tool for helping with legal procedures for lawyers, mediators and judges, with the objective of reaching an agreement between the parties''. The present work describes in detail one of the procedures developed exclusively by the author in the context of this project, with the valuable feedback of the participants in the ``De Aequa Divisione Workshop'' within the project.

The theory of fair division dates back to the end of the Second World War. It was devised by a group of Polish mathematicians, Hugo Steinhaus, Bronisław Knaster and Stefan Banach, who used to meet in the Scottish Caf\'e in Lvov  (see Knaster \cite{k46}, Steinhaus\cite{s48} and \cite{s49}). For an account of the many results that followed, I refer to the books by Brams and Taylor \cite{bt96}, Moulin \cite{m03} and to the review papers of Bouveret et al.\ \cite{bcm16}, Moulin \cite{m19}, Procaccia \cite{p16}  and Thomson \cite{t11} and \cite{t16}.
 
Many methods rely on point allocation methods. Quoting its Wikipedia page, the Adjusted Winner procedure, one of the most popular fair division procedures proposed by Brams and Taylor \cite{bt96} explains that ``each player is given the list of goods and an equal number of points to distribute among them. He or she assigns a value to each good and submits it sealed to an arbiter.''
In an actual division of goods, however, the role of money and the market value of the disputed items involved cannot be ignored. This happens for several reasons:
\begin{enumerate}[i)]
\item The law in many countries requires that the allocation of goods in an inheritance, a divorce or a company liquidation yields bundles of equal  market value (or proportional to each agent's entitlement). No reference is usually made to the agents' satisfaction. 
\item Money itself plays a role in the division of goods. This can happen for several reasons:
\begin{enumerate}
\item Money can be used to level out the disparities arising in the distribution of the other goods, if they are indivisible, or if they are not perfectly divisible\footnote{It is well known that when all the preferences are represented by non-atomic measures, fair allocations do exist. See, for instance, \cite{ds61} and \cite{w85}. } (because they contain indivisible ``lumps''). In a similar manner, if a good has to be split among several agents, but the division is not practical or it is unsatisfactory, an agent may buy the shares of the other agents involved and become the sole owner. 
\item In the search for a better solution, one or more goods may be sold to external agents, and the resulting cash would replace the item in the division. A model that makes explicit reference to this selling option, and to the comparison of the agents' utilities with a market value in the two-agent case is provided by Karp et al. \cite{kkp14}.
\item The previous case is just one of the many situations where money is one of the goods to be divided. Early examples of models in which money and a set of indivisible goods are allocated are given by Alkan et al.\ \cite{adg91} and Bevia \cite{b98}.
\end{enumerate}
In all the examples above, agents need to compare the satisfaction of each good with that of money -- and,  therefore, they define a personal monetary evaluation for  each contended good. 
\item Even when money is not explicitly evoked, asking the agents to assign a monetary value to each good is a reasonable method for eliciting their preferences. 

 \end{enumerate}
 
 How can we take account of the agents' satisfaction?
A first answer has been proposed by Bellucci and Zeleznikow \cite{bz06}, with the Family Winner procedure, which was later perfected by Bellucci in \cite{b08} and by Abrahams et al.\ in \cite{abz12} as the Asset Divider procedure. The procedure combines the goods' market values and ratings, which are then turned into a fixed number of points allocated among the goods, in the spirit of Brams and Taylor's Adjusted Winner procedure. The output is an allocation that is equal in the market values of the bundles. In the cited references, no formal code is given, but only a verbal description of the procedure. The Asset Divider  allocates one good at a time, starting from the goods with highest number of points for each side, and after each transfer modifies the preferences of the goods still unassigned. It proceeds until the dollar amount of the bundles does not exceed the percentage split indicated by the mediators. Goods are indivisible, but the procedure may indicate a money transfer to make the division fairer.
The optimality properties of the suggested allocation are not examined in detail as they do not fall within the scope of the work. 
As a  consequence of this mathematical indeterminacy, I note from the examples that illustrate how the Asset Divider procedure works that, while the bundles of the two agents are equal in their monetary value, they may differ in terms of utility, with no explanation provided. The procedure deals with the two-agent case, with no easy generalization for a larger group of agents.
 
 Another contribution that deals with the two-agent case is the above mentioned paper by Karp et al.\ \cite{kkp14}. The authors examine the boost in terms of efficiency that results from introducing the option of selling some of the contended goods and distributing the resulting cash -- instead of allocating that good to one of the two parties. Utilities and monetary values are measured on the same scale. Therefore, the utility is the willingness to pay for a given item, while the selling price is defined as a constant fraction of the minimum between the two agents' evaluations. The most recent addition is provided by Bogomolnaia and Moulin \cite{bm22}. Here, two new procedures are designed to allocate indivisible goods with cash transfers. In Sell\&Buy, agents bid for the role of Seller or Buyer: with two agents, the smallest bid defines the Seller who then charges a price constrained only by their winning bid.
In Divide\&Choose, agents bid for the role of Divider, then everyone
bids on the shares of the Divider’s partition.
 
 In the European project that financed the present work, another procedure was designed in which agents are asked to place bids from a virtual personal budget. The agent's utility is set equal to the bid and the Nash/Competitive solution is computed. More details about this other procedure can be found in \cite{d19}.

 \subsection{A New Proposal}
 I recognize the importance of dealing with two distinct sources of information: the market values of goods and the agents' preferences, defined here as the willingness to receive a good. The approach that I adopt, however, is quite different from that of Bellucci \cite{b08}. First of all, I consider divisible goods, instead of indivisible ones, and I devise a procedure that works for any number of agents. 
 
 More importantly, though, the goal is to overcome the difficulties highlighted above, and frame the problem against a solid mathematical background.
 The first challenge is to find a proper way to model the agents' preferences when both the market values and the agents' likes and dislikes must be taken into account. There is a consolidated stream of literature on how to represent preferences for resource allocation problems. Preferences can be cardinal (i.e.\  utilities) or ordinal (rankings, typically), they may regard single goods valued independently, or they may refer to bundles -- thus underlining the dependencies that may occur upon the goods' reception. A recent review of the different approaches is given by Bouveret et al.\ \cite{bcm16}.
 
  A general underlying principle holds: the quality of information behind each of these methods varies, and the higher the quality, the more difficult it is to elicit the corresponding piece of information from the agents. For instance, it is easier for the agents to evaluate single goods, rather than all the possible bundles or a large part of them, even if this means losing some information about the dependencies among the goods. I stick to the simpler settings and assume preferences to be separable or additive. With regard to single goods, the market value is a cardinal piece of information with high informative value: assuming a positive dependence between the agents' preferences and the goods' market value,   I  propose a framework where personal taste affects the market value by modifying it. The value distortions take place by means of a simple rating feedback: agents are asked to rate each good on a fixed range with an odd number of levels -- typically 5. The median rating will make the market value and the utility coincide, while higher (lower, resp.) will increase (decrease, resp.) the utility by a constant factor applied a number of times given by the distance from the median rating.
 
 Choosing a small range of levels (5, or even 3) may lead to a utility which is an approximation of the exact measure for the agent's fondness for a given item. The emphasis is, however, on the simplicity of the elicitation process. Once the agents communicate their ratings for each good, their task is completed. Moreover,  even if the introduction of a discrete rating range may suggest an intermediate level between cardinal and ordinal preferences, the quality of information provided by market values is too valuable to be downgraded as ordinal, and the market value modification mechanism is considered instead. Once the market value of goods and the agents' preferences are put together into a single utility function, linear programming is used to return the solution of an optimization problem, in stark contrast to the Asset Divider's approach.
 
By setting an optimality criterion properly, I address (and overcome) a difficulty arising from the fact that optimally fair divisions that provide equal (or proportional) market shares may fail to exist, while imposing both criteria may not only produce outcomes which are unacceptable for the agents than those resulting from arrangements with unequal market shares, but may also return too many split items. The proposed solution drops the requirement of exactly equal (or proportional)  market value shares, but fully justifies the differences that may arise in terms of different average ratings of the goods per monetary unit received: if an agent receives a larger market value share than that of another other agent, they will receive  goods with different average satisfaction per monetary share (smaller for the agent who got a larger market share). A rigorous framework will be provided to measure those discrepancies. As an additional benefit, the procedure always guarantees the minimal number of split items, which is given by the number of agents involved in the division minus 1. 

Most of the known available procedures, including the Adjusted Winner and the Asset Divider procedures, enjoy the property of scale invariance: if all the ratings by a player are multiplied by a constant, the global outcome is unchanged. While sound in principle, this property has little cogency in many situations where a ``scaled'' profile is not suitable. Consider the case when utility is based on a discrete scale, such as when an agent allocates an integer number of points to each good. In this situation, it is therefore often impossible to obtain a new profile from an old one by using a simple scalar multiplication (especially when the scalar is less than 1) and it is impossible to compare different profiles by invoking scale invariance. I consider a different property, translation invariance, in which, when all ratings are increased, or decreased by the same amount, the outcome is unchanged. This property holds whenever the adopted utility model is applied to a scale invariant solution, and has proved easy to convey to audiences of law researchers and practitioners.

\section{Notation and Assumptions}
I consider a (finite) set of $g$ valuable items $G$ (which may also be referred to as {\em goods}) to be divided among a (finite) set $N$ of $n$ agents. The share of a good $a$ assigned to agent $i$ is denoted as $z_{ia} \in [0,1]$. For each $i \in N$, vector 
\[
z_i=\begin{bmatrix}
z_{i1}
\\
z_{i2}
\\
\vdots
\\
z_{ig}
\end{bmatrix} \in [0,1]^G
\]
denotes agent $i$'s allocation. The entire allocation is grouped as a matrix
 $z=\left( z_1,\ldots,z_n\right)$. We assume that only one unit per good is distributed and that, since all goods have a market value, only allocations that assign the entire good to one or more agents are considered (this is often referred to as  the ``nonwastefulness'' hypothesis). These allocations are defined {\em feasible} and the set of such allocations is denoted as 
 \[
 \Phi(N,G) =\left\{  z: \sum_i z_{ia} =1 \mbox{ for every } a \in G \right\} \; .
 \] 
Typically, no pair of agents values the same good equally. The degree of appreciation for good $a \in G$ by agent $i \in N$ is described by a non-negative number $u_{ia}$, conveniently arranged in vectors $u_i =(u_{i1}\ldots,u_{ig})^T \in \mathbb{R}_+^G$ and then in a matrix  $u=(u_1,\ldots,u_n)$. A division problem is then fully characterized by the triplet $\mathcal{Q}=(N,G,u)$. I now assume that utilities are {\em additive} (the utility of receiving a bundle of goods equals the sum of the utilities of the single goods) and {\em linear} (the utility of receiving the fraction of a good equals the same fraction of the utility for receiving the good in its entirety). Consequently, an allocation $z \in \Phi(N,G)$ will give a utility of $U_i(z)=u_i \cdot z_i=\sum_a z_{ia} u_{ia}$ for agent $i \in N$. Let $U(z)=(U_1(z),\ldots,U_n(z))^T$ be the utility profile corresponding to an allocation and let $IPS(N,G,u)$ (or simply $IPS$ if the context is clear\footnote{$IPS$ is an acronym for {\em Individual Pieces Set}, as given in Definition 4.1 in Barbanel \cite{b05}}) be the set of utility profiles corresponding to feasible allocations. Notice that the utility of receiving the set of all goods need not be the same for all the agents, i.e.\ utilities are not normalized. 

All agents are usually assumed to have the same importance in the allocation. In practical situations, agents may have different weights that measure their importance. Consider, for instance, the different degrees of kinship in an inheritance case, or the different contribution to a marriage that might induce a judge to assign a larger share of the joint assets to one of the ex-spouses. Let $w=(w_1,\ldots,w_n)^T$ be the vector of agents' weights, with $w_i>0$ for every $i \in N$ and $\sum_{i \in N} w_i=1$. When weights are not mentioned, it is assumed that $w_i=n^{-1}$ for every $i \in N$.

\subsection{Measuring fairness} 
What makes an allocation fair? The question does not come with an easy answer, and this is what makes the whole topic interesting. The proposed allocation often stands out as a solution for an objective function which measures the  agents' global satisfaction. Among the many proposals, two functions convey an idea of fairness particularly well, and have resurfaced over and over again in the specific literature for theory and applications: 
\begin{itemize}
\item The Egalitarian solution (which is derived from the Egalitarian Equivalent allocation by Pazner and Schmeidler \cite{ps78}) defined as 
\begin{equation}
\label{egsol}
z^e\in 
\argmax_{z \in \Phi(N,G)} \min \frac{\bar{U}_i(z)}{w_i},
\end{equation}
where $\bar{U}_i(z)=\frac{U_i(z)}{\sum_a u_{ia}}$ is the {\em normalized} utility for agent $i \in N$. The solution name comes from the fact that, under mild conditions such as the requirement that every good has some positive value for every agent, i.e.\ $u_{ia}>0$ for every $i \in N$ and $a \in G$, it turns out (see Corollary 5.8 in  Dall'Aglio \cite{d01}) that the solution guarantees an equal and normalized utility for all agents, namely
\begin{equation}
\label{egaldef}
\frac{\bar{U}_i(z)}{w_i}=\frac{\bar{U}_j(z)}{w_j} \qquad \mbox{for every } i,j\in N
\end{equation}
\item The Nash/Competitive solution (after Nash \cite{n50}) defined as
\begin{equation}
\label{egsol}
z^c\in 
\argmax_{z \in \Phi(N,G)} \prod U_i(z)^{w_i}
\end{equation}
Nash introduced this solution in the context of bargaining problems. The solution is also referred to as ``competitive'' because it can be obtained as a competitive equilibrium in the exchange economy, where each
agent is endowed with an amount of money proportional to the agent's weight (see Bogomolnaia et al.\ \cite{bmsy17a} and \cite{bmsy17b} for an updated review of these results). 
\end{itemize}
As may be expected, no criterion prevails over the other. First of all, both solutions satisfy the following {\em invariance by scale property}: suppose that the utility profile of each agent is multiplied (scaled) by some constant $\lambda_i > 0$ for every $i \in N$, i.e.,
\[
u'_{ia} = \lambda_i u_{ia} \qquad \mbox{for every } i \in N \mbox{ and } a \in G,
\]
then $\mathcal{Q}=(N,G,u)$ and $\mathcal{Q}'=(N,G,u')$ yield the same solution sets.

In the case of the Egalitarian solution, this is true because normalized utilities in the objective function are considered. For the Nash/Competitive one, $U'_i(z)=\lambda_i U_i(z)$, and thus, since a logarithm appears in the objective functions, the two problems $\mathcal{Q}$ and $\mathcal{Q}'$ simply differ by a constant and are equivalent.

The Egalitarian and the Nash/Competitive solutions share other 
important properties, such as:
\begin{itemize}
\item Fair Share Guarantee: everyone is guaranteed at least their fair share of all of the assets. In formula:
\[
U_i(z) \geq w_i \sum_a u_{ia} \qquad \mbox{for every } i \in N
\]
\item Efficiency: The allocation implements an efficient utility profile, a vector $U$ of utility values that is not Pareto dominated, i.e.\ it cannot be improved upon by another allocation that assigns greater utility to at least one of the agents. In formula:
\[
U(z) \leq U'(z) \mbox{ and } U'(z) \in IPS(N,G,u) \Longrightarrow 
U(z)=U'(z)
\]
\end{itemize}
The two solutions differ in the fulfilment of other important properties: as already stated, the Egalitarian solution, under mild conditions, satisfies \eqref{egaldef}, while the Nash/Competitive one typically fails the same test. Conversely, the latter always satisfies 
\begin{itemize}
\item No Envy: $\frac{u_i \cdot z_i}{w_i} \geq \frac{u_i \cdot z_j}{w_j}$ for all $i$, $j$.

In other words, every agent values the received bundle of items at least as much as the bundles assigned to the other agents.
\end{itemize}
The Egalitarian solution satisfies the same property for the two-agent case, but may fail to do so when the number of agents increases to 3 or more (see p.84 in Brams and Taylor \cite{bt96} and Theorem 3.10 in Dall'Aglio and Hill \cite{dh03}).

The comparison may continue and I refer to Bogomolnaia et al.\ \cite{bmsy17a},\cite{bmsy17b} and Moulin  \cite{m19} for a recent and thorough comparison of the two solutions in the linear setting adopted here, and in more general frameworks.  My choice for the procedure that I am going to illustrate will not be based on an a priori opinion, but will be justified by the properties that it guarantees, given the assumptions and the data available for any instance.

\section{A Procedure With Market Values with Preferences}
The utility for any agent of any good is expressed as a positive number. Since in the present model, goods have a market value, their utility should be comparable to this value. Agents' inclinations may make the goods more (or less) valuable than their face value, but, in any case, the good's utility for any agent should not be too distant from the good's monetary value, because valuable goods can usually be traded outside the division context. Under these assumptions, it is reasonable to put bounds on the utility of a good by defining an interval of reasonable values that includes the market value. 

Economics models take great care in the mathematical description of the agents' utilities, but often overlook the process of eliciting the agents' preferences inside the model. 
Eliciting preferences typically requires a careful balance between precision in the specification of the agents' preferences and simplicity in the definition of rules that can be understood  by  a vast audience of non-specialists, with little or no background in mathematics or economics. 
The method that I propose supports the latter feature, while maintaining, under some proper assumptions, a sufficient degree of the former. It uses an old intuitive method, which is experiencing a resurgence in popularity: a rating system induced by the repetition of a given symbol, typically  a star. The method dates back to an 1820 guidebook by Mariana Starke \cite{s1820}. Since then, it has been used by critics to grade artworks (books, movies, theatrical performances, and so on) or by travellers or institutions to evaluate facilities (hotels, restaurants, and so on). More recently, it has become a popular method used by major internet goods and services retailers (such as Amazon, eBay or TripAdvisor) to let customers provide feedback on their consumption experience for other customers to make more informed decisions. Its popularity should put every potential user of the procedure at ease. 
The rating system selects a finite number of values from the continuous range of plausible choices. 

\begin{definition}
The {\em Discrete Power Rating (DPR)} utility model on goods with market values $m=(m_1,\ldots,m_g)^T$ in a rating scale with $2q+1$ levels is defined as
\begin{equation}
\label{ratuti}
u_{ia}=K^{r_{ia}-q-1} \cdot m_a \qquad \mbox{for each } i \in N  \mbox{ and each } a \in G
\end{equation}
where $m_a$ is the market value of good $a$, $r_{ia} \in \{1,2,\ldots,2q+1\}$ is the rating by agent $i$ of good $a$ and $K>1$ is the constant multiplicative factor.
\end{definition}

 Note that the range of plausible values for a good's utility is given by the interval $[K^{-q}m_a,K^{q}m_a]$, and $2q+1$ values are selected in the interval. The median rating $q+1$ will yield $u_{ia}=m_a$, while higher (lower, resp.) ratings will increase (decrease, resp.) the utility by the constant factor $K$ ($K^{-1}$, resp), applied a number of times given by the distance from the median rating.

The discretization process defines  a finite grid of values that the utility may assume. Moving from a continuous range to a discrete approximation may cause a distortion of the agents' utlities. Clearly the grid may be dense in the real numbers by means of a proper choice of the parameters $K$ and $q$, but this would aggravate the elicitation process. For instance, the average user may be unable to distinguish between ratings 22 and 23 in a range of 87 levels.  The robustness study that follows supports the use of a small grid.  A rating system with 5 levels, i.e.\ $q=2$, is rich enough for most purposes: it encompasses the 3-level system by considering a restricted grid in which only ratings of 5 stars, 3 stars or 1 star are permitted). I will use this range for all the examples and the applications that follow. 

Giving the maximum rating for a good increases the chances of receiving that good, but it does not grant any of it. This happens because other agents may give maximum ratings too and, more generally, because fairness requirements may mean allocating the goods in a different way.

When agents rate the goods, a simple invariance principle could guide the agents in revealing their genuine preferences.  If agents understand it, 
they should assign high ratings only to the goods they really care about, rather than assign a lot of stars in the vain hope of receiving a richer share  than that of the others. Scale invariance, the most popular among these notions, is hard to put into practice. Each agent should be able to scale up or down a profile of preferences by a constant factor in order to be able to compare different sets of utilities. In the case of a discrete range of ratings and, a fortiori, with such a limited number of levels, the principle becomes impossible to implement. 

Instead of scale invariance, the DPR model applied to a scale invariant solution satisfies the following:
\vskip0.3cm
\noindent {\em Translation invariance}: Adding or removing one star to the ratings of all of the goods will not change the outcome. 
\vskip0.3cm

The principle applies to fractions of stars as well, if such ratings are allowed by the system.
 To exemplify things in a 5-level rating scale, according to the translation invariance principle, the profiles in Table \ref  {1stinst_rat} yield the same outcome.
 
 \begin{table}[h!]
\begin{center}
\caption{First instance of translation invariance}
\label{1stinst_rat}
 \begin{tabular}{||c| |c| c|c ||} 
 \hline
 Items & Profile A & Profile B &  Profile C \\
 \hline\hline
Town H.\ & \mstab & \nstab & \pstab 
\\ 
 \hline
Country H.\ & \nstab & \pstab & \ppstab 
\\ 
 \hline
Car & \mmstab & \mstab & \nstab 
\\ 
 \hline
M. bike & \nstab & \pstab & \ppstab 
\\ 
 \hline
 Garage & \mstab & \nstab & \pstab 
\\ 
 \hline
\end{tabular}
\end{center}
\end{table}
Considering a more extreme case,  all of the profiles described in Table \ref{2ndinst_rat} indicate indifference towards the goods.
\begin{table}[h!]
\begin{center}
\caption{Second instance of translation invariance}
\label{2ndinst_rat}
 \begin{tabular}{||c| |c| c|c ||} 
 \hline
 Items & Profile D & Profile E &  Profile F \\
 \hline\hline
Town H.\ & \mmstab & \nstab & \ppstab 
\\ 
 \hline
Country H.\ & \mmstab & \nstab & \ppstab 
\\ 
 \hline
Car &\mmstab & \nstab & \ppstab
\\ 
 \hline
M. bike & \mmstab & \nstab & \ppstab 
\\ 
 \hline
 Garage & \mmstab & \nstab & \ppstab 
\\ 
 \hline
\end{tabular}
\end{center}
\end{table}
In other words, adding too many stars will result in an expression of indifference among goods.
What counts, instead, is a correct profiling, where the agents are able to indicate which goods they really care about.

\subsection{Choosing the right objective}
When monetary values are explicitly defined, the worth of what each player gets cannot be ignored.
\begin{definition} The {\em market} or {\em monetary value} $\mu_i(z)$ of the bundle received by agent $i \in N$ is defined as
\begin{equation}
\label{mv_bundle}
\mu_i(z) = \sum_{a \in A} m_a z_{ia} \qquad i \in N .
\end{equation}
\end{definition}

In the present model, I look for an allocation with the following properties: $(i)$ it is fair according to an established criterion and $(ii)$ it provides the agents with bundles of equal market values (or proportional to the entitlement shares). According to the short review in Section 2.1, the Egalitarian or the Nash/Competitive solutions are good candidates for property $(i)$, while property $(ii)$ can be taken care of by considering a restricted class of allocations.

\begin{definition}
 Let $M=\sum_{a \in A} m_a$ be the goods' total value and consider agents with entitlements given by the vector $w$. A feasible allocation $z$ is Proportional in the Market Values (PMV) if
 \begin{equation}
\label{def:pmv}
\mu_i(z)=M w_i  \qquad \mbox{for every } i \in N.
\end{equation}
Let $\Phi_{PMV}(N,G,m,w)$ denote the set of PMV allocations for the allocation of divisible goods in $G$ with market prices $m$ to agents in $N$ with entitlements $w$.
\end{definition}

It is therefore natural to consider the following optimal allocations.

\begin{definition}
A Proportional Egalitarian solution $z^e_{PMV}$ is defined as
\[
z^e_{PMV}\in 
\argmax_{z \in \Phi_{PMV}(N,G,m,w)} \min \frac{\bar{U}_i(z)}{w_i}.
\]
Similarly, a proporional Nash/Competitive solution $z^c_{PMV}$ is defined as
\[
z^c_{PMV}\in 
\argmax_{z \in \Phi_{PMV}(N,G,m,w)} \prod U_i(z)^{w_i}.
\]
\end{definition}

The following simple example shows why such a natural proposal turns out to be a bad idea.

\begin{xmpl}
\label{xmpl1}
Consider two agents with equal importance: $I$ and $II$ who divide 2 items $A$ and $B$ of equal market value between themselves. The items' market values, together with the agents' ratings are shown in Table \ref{ratings_toy}, with $K >1$.

\begin{table}[h!]
\begin{center}
\caption{Data for Example \ref{xmpl1}.}
\label{ratings_toy}
 \begin{tabular}{||c | c || c| c||c|c ||} 
 \hline
 Item & mv & $r_I$ & $u_I$ & $r_{II}$ & $u_{II}$ \\  
 \hline\hline
$A$ & $1$ & \nstab & $1$  & \mstab & $K^{-1}$ 
\\ 
 \hline
$B$ & $1$ & \pstab & $K$  & \ppstab & $K^2$ 
\\ 
 \hline \hline
 \end{tabular}
\end{center}
\end{table}

To compute the (unconstrained) Egalitarian allocation, note that the allocation is efficient, and equality in the utility is reached only when item $A$ is assigned to agent 1, while $B$ is split between the two. Therefore, the allocation must be sought in 
\[
\Phi_{B}=  \left\{  z(b)= \begin{bmatrix}
1 & 1-b\\ 0 & b
\end{bmatrix}: 0 \leq b \leq 1  \right\}
\]
Some algebra is required to show that agents get equal utility when $b^e=(K^3+1)/(2K^3-K^2+K)$. Therefore 
\begin{gather}
\label{xmpl:allocegal}
z^{e}=
\kbordermatrix{
     & $A$ & $B$ 
     \\
     I & 1 & \frac{K^3-K^2+K-1}{K(2K^2-K+1)} 
     \\
     II & 0 & \frac{K^3+1}{K(2K^2-K+1)} }\quad \mbox{ and}
\\
\label{xmpl:utilegal}
\bar{U}_1(z^e)=\bar{U}_2(z^e)= \frac{K^2}{2K^2-K+1} > \frac{1}{2}
\end{gather}
It can be shown that splitting item $A$ yields a product of utilitities lower than that obtained by splitting item $B$. Therefore, the unconstrained Nash/Competitive allocation must also be sought in $\Phi_B$. This product is maximal when $b^c=\frac{1+K}{2K}$. Therefore
\begin{gather*}
z^{c}=
\kbordermatrix{
     & $A$ & $B$ 
     \\
     I & 1 & \frac{K-1}{2K} 
     \\
     II & 0 & \frac{K+1}{2K} },
\\
\bar{U}_1(z^c)=\frac{1}{2} \quad  \mbox{ and} \quad  \bar{U}_2(z^c)=\frac{K^2}{2(K^2-K+1)} > \frac{1}{2}
\end{gather*}

The PMV allocations when $w=(1/2,1/2)^T$ and $m=(1,1)^T$ are 
\[
\Phi_{PMV}(m,w)= \left\{ z(a)=\begin{bmatrix}
a & 1-a \\ 1-a & a
\end{bmatrix}: 0 \leq a \leq 1  \right\}.
\]
Equality between utility is obtained when $a=1/2$. Thus,
\[
z^{e}_{PMV}=
\kbordermatrix{
     & $A$ & $B$ 
     \\
     I & 1/2 & 1/2 
     \\
     II & 1/2 & 1/2 } \mbox{ and }
\bar{U}_1(z^e_{PMV})=\bar{U}_2(z^e_{PMV})=\frac{1}{2} 
\]
This solution is strongly dominated by the unconstrained Egalitarian solution -- and it is therefore inefficient. 

Moving to the Nash solution among PMV allocations, the product $\bar{U}_1(z(a))\bar{U}_2(z(a))$ is quadratic in $a$ and achieves its maximum in $(K^3+K^2+K-1)/(2(K^3-1))$. Since $a$ is constrained in $[0,1]$, the maximal product is achieved by
\[
a^c=\begin{cases}
1 & \mbox{if }1 < K \leq K_0
\\
\frac{K^3+K^2+K-1}{2(K^3-1)} & \mbox{if }K > K_0
\end{cases}
\]
where $K_0 \approx 1.839$ is the only real root of $K^3-K^2-K-1$. Thus,
\[
z^c_{PMV}=\left\{
\begin{array}{cl}
\begin{bmatrix}
1 & 0 
     \\
0 & 1 
\end{bmatrix} & \mbox{ if } 1< K \leq K_0  
\\
 & 
 \\
\begin{bmatrix} \frac{K^3+K^2+K-1}{2(K^3-1)} &  \frac{K^3-K^2-K-1}{2(K^3-1)} 
     \\
\frac{K^3-K^2-K-1}{2(K^3-1)} & \frac{K^3+K^2+K-1}{2(K^3-1)}   
\end{bmatrix} & \mbox{ if } K > K_0
\end{array}
\right.
\]
The normalized utilities for the allocation are
\begin{gather*}
\bar{U}_1(z^c_{PMV})= \begin{cases}
\frac{1}{K+1} & \mbox{if } 1 < K \leq K_0
\\
\frac{K^2+1}{2(K^2+K+1)} & \mbox{if } K > K_0
\end{cases} \mbox{ and } 
\\
\bar{U}_2(z^c_{PMV})= \begin{cases}
\frac{K^3}{K^3+1} & \mbox{if } 1 < K \leq K_0
\\
\frac{K^2+1}{2(K^2-K+1)} & \mbox{if } K > K_0
\end{cases}
\end{gather*}
It can be easily verified that $\bar{U}_1(z^c_{PMV})< 1/2$ for any $K>1$ and, therefore, the Proportional Nash/Competitve solution fails the Fair-Share Guarantee.

Figure \ref{fig:xmpl1} shows the $IPS$ with normalized utilities in the example when $K=1.2$. 
\begin{figure}
     \centering
     \begin{subfigure}[b]{0.7\textwidth}
         \centering
         \includegraphics[width=0.8\textwidth]{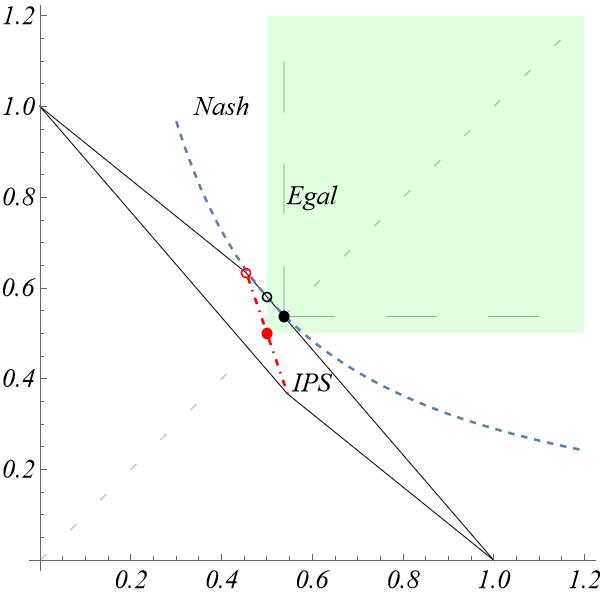}
         \caption{The Individual Pieces Set (IPS)}
         \label{fig:xmpl1IPS}
     \end{subfigure}
     \hfill
     \begin{subfigure}[b]{0.25\textwidth}
         \centering
         \includegraphics[width=0.8\textwidth]{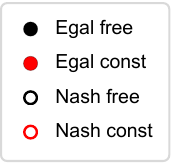}
         \caption{Legend}
         \label{fig:xmpl1Legend}
     \end{subfigure}
        \caption{A geometrical description of Example \ref{xmpl1} when $K=1.2$. The green area indicates allocations that satisfy the fair share guarantee.}
        \label{fig:xmpl1}
\end{figure}
\end{xmpl}

The above example shows an instance where the Proportional Egalitarian or the Proportional Nash/Competitive solutions do not meet the minimal requirements for being considered acceptable. In fact, the Proportional Egalitarian solution is inefficient, while the Proportional Nash/Competitive one does not satisfy the Fair Share Guarantee.

There is another important disadvantage that arises when equal market values are imposed: whether optimal or not, the solution may require a larger number of items to be split. Proposing a solution with many split items poses many practical challenges: When an item cannot actually split, an arrangement has to be found among the agents who receive a portion of the good to either manage the good together or find a satisfactory resolution. The larger the number of split items, the more difficult it is to achieve a peaceful outcome and the concern is especially important if the total number of items is small. In Example \ref{xmpl1}, all the allocations which are proportional in the market values and satisfy the Fair Share Guarantee require both the items to be split. This fact must be compared with a recent result regarding the division of homogeneous and divisible goods. Lemma 2.5 in Sandomirskiy and Segal-Halevi \cite{ss19} state that any Pareto efficient profile deriving from the division of goods among $n$ agents, including those corresponding to the Egalitarian or the Nash/Competitive solutions, can be obtained by splitting at most $n-1$ goods. 
Therefore, the pursuit of an allocation that is Proportional in the Market Values and satisfies the Fair Share guarantee may conflict with a further goal:  keeping the number of split items to a minimum. 

\subsection{The Egalitarian solution for the DPR utility model}
\label{subs_procedure}
Example \ref{xmpl1}  provides strong arguments for rejecting any procedure aimed at satisfying fairness and perfect equality in market values. Instead of imposing new requirements for the solution, I return to the unconstrained model. The well-known solutions (either the Egalitarian or the Nash/Competitive) will not, in general, return bundles of equal market value -- as pointed out by Example \ref{xmpl1}. In my proposal, I will let the bundles diverge in their market value, but I will find a key to explain those differences. I begin by applying the known solutions to an example. For the computations in this example and the ones that follow, I used the Mathematica 12.3 software.

\begin{xmpl} 
\label{xmpl:toy_Moulin}
Consider a situation with 3 goods ($A$, $B$, $C$) and 4 agents ($I$, $II$, $III$, $IV$)  that evaluate the goods on a 5-level rating scale. The goods' values and the agents' ratings are given in Table \ref{rat_toy_Moulin}.

\begin{table}[h!]
\begin{center}
\caption{The ratings and values for Example \ref{xmpl:toy_Moulin}.}
\label{rat_toy_Moulin}
\begin{tabular}{||c|c||c|c|c|c||}
\hline \hline
Item & Market v.\ & $I$ & $II$ & $III$ & $IV$\\
\hline \hline
$A$ & 100 & \ppstab  & \mmstab& \mstab & \nstab \\
\hline
$B$ & 200 & \mstab  & \ppstab & \mmstab & \nstab \\
\hline
$C$ & 300 & \mmstab  & \mstab & \ppstab & \nstab\\
\hline
\end{tabular}
\end{center}
\end{table}

For values of $K$ close to 1, the Egalitarian and the Nash/Competitive allocations do not differ too much in their structure, since the goods or their fractions are allocated to the same agents. For instance, if we set $K=1.1$, we obtain the Egalitarian and the Nash/Competitive, denoted as  $z^e$ and $z^c$, respectively, as
\[
z^{e}=
\kbordermatrix{
     & A & B & C
     \\
     I & 1 & 0 & 0.1426  
     \\
     II & 0 & 0.6865 & 0
     \\
     III & 0 & 0 & 0.4944  
     \\
     IV & 0 & 0.3135 & 0.3630
} \quad
z^{c}=
\kbordermatrix{
     & A & B & C
     \\
     I & 1 & 0 & 0.0839  
     \\
     II & 0 & 0.7914 & 0
     \\
     III & 0 & 0 & 0.5276
     \\
     IV & 0 & 0.2086 & 0.3885
}
\]
As expected, the monetary values of the bundles received by the agents differ in the two solutions. In the Egalitarian solution, these values are all different, while three out of four values coincide in the Nash/Competitive solution, as shown in Table \ref{monval_toy_Moulin}.
\begin{table}[h!]
\begin{center}
\caption{The monetary values for $K=1.1$ in Example \ref{xmpl:toy_Moulin}.}
\label{monval_toy_Moulin}
\begin{tabular}{||c||c|c||}
\hline \hline
Agent & Egalitarian & Nash/Competitive\\
\hline \hline
$I$ & 142.78 & 125.17 \\
\hline
$II$ & 137.30 & 158.27 \\
\hline
$III$ & 148.32 & 158.27 \\
\hline
$IV$ & 171.61 & 158.27 \\
\hline
\end{tabular}
\end{center}
\end{table}
The monetary values of the agents' bundles as $K$ spans the interval $[1,1.5]$, are plotted in Figure \ref{fig:monval_Moulin}. 
\begin{figure}
     \centering
     \begin{subfigure}[b]{0.45\textwidth}
         \centering
         \includegraphics[width=\textwidth]{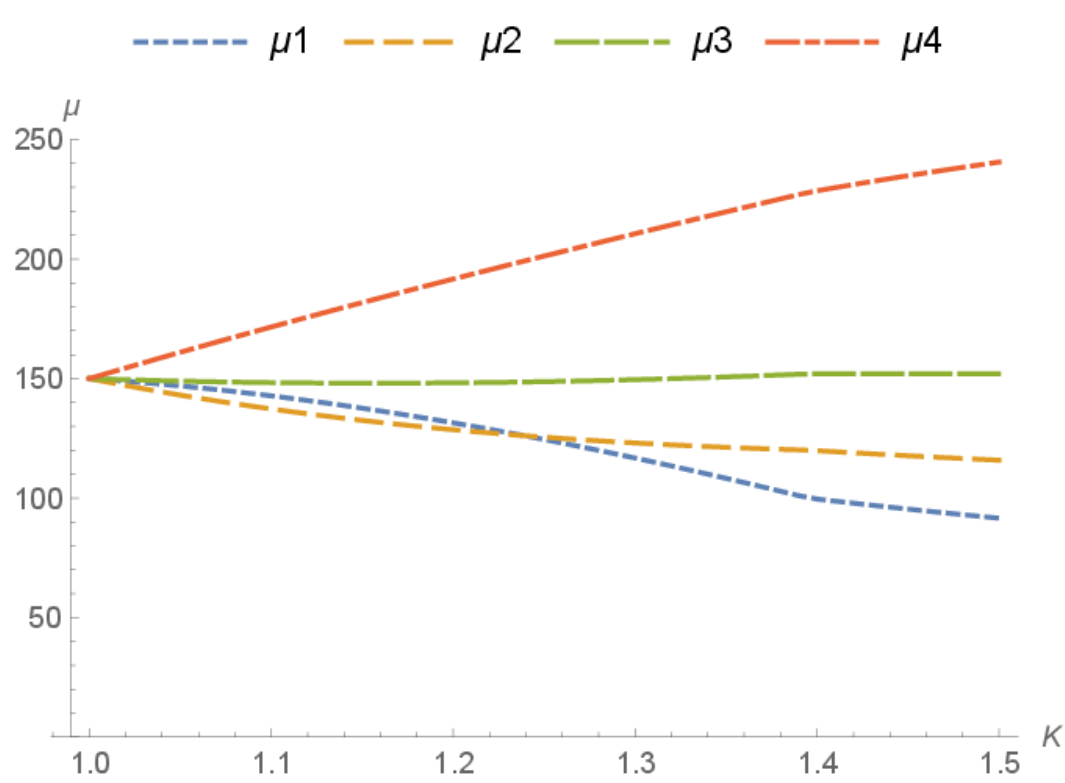}
         \caption{The Egalitarian solution}
         \label{fig:valmoneq}
     \end{subfigure}
     \hfill
     \begin{subfigure}[b]{0.45\textwidth}
         \centering
         \includegraphics[width=\textwidth]{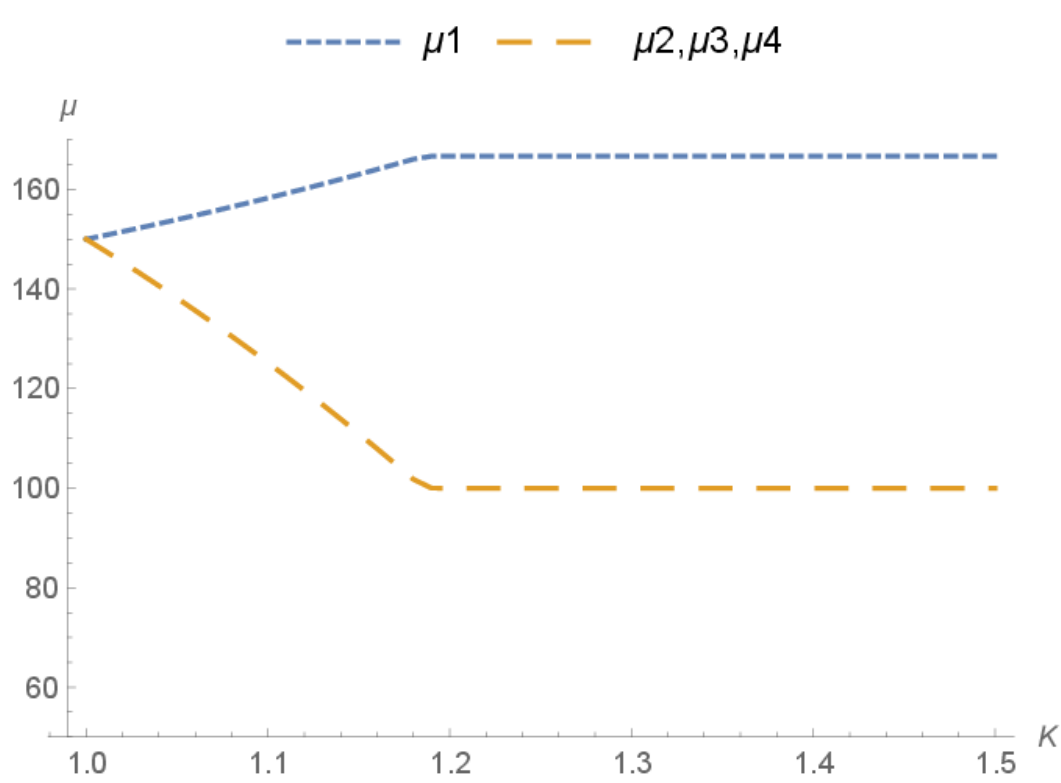}
         \caption{The Nash/Competitive solution}
         \label{fig:valmoncp}
     \end{subfigure}
        \caption{Monetary values in Example \ref{xmpl:toy_Moulin}.}
        \label{fig:monval_Moulin}
\end{figure}
We notice that the market values for the Egalitarian solution are all different and keep spreading as $K$ increases. Furthermore, the graphs of the market values for agents $I$ and $II$ cross at $K \approx 1.2391$. For the Nash/Competitive solution, the gap between the common monetary value of agents $II$, $III$ and $IV$, and that of agent $I$ keeps increasing up to the value $K^*=\sqrt[3]{5/3} \approx 1.18563$. For values of $K$ larger than or equal to that threshold, the Nash/Competitive partition remains constant and equal to
\[
z^{c}=
\kbordermatrix{
     & A & B & C
     \\
     $I$ & 1 & 0 & 0  
     \\
     $II$ & 0 & 5/6 & 0
     \\
     $III$ & 0 & 0 & 5/9
     \\
     $IV$ & 0 & 1/6 & 4/9
}.
\]
This fact explains the flat part in the graph in Figure \ref{fig:valmoncp}.
\end{xmpl}
In what follows, I am going to provide an interpretative key to the Egalitarian solution: differences in the market value of the bundles will be compensated by the average satisfaction that each fraction of monetary value associated with the goods will provide to the agent. For this reason, I propose an allocation which is Egalitarian in the agents' normalized utilities. 
In making this choice, I do not claim that the Egalitarian solution is better than the Nash/Competitive one.  I simply could not find a similar justification for the market values attributed by the Nash/Competitive solution. In Example \ref{xmpl:toy_Moulin}, why does the market value of agent $I$ differ from that of the other three agents, and why do the values for agents $II$, $III$ and $IV$ coincide? 

I now describe a procedure for obtaining the Egalitarian solution.
\vskip0.3cm

{\bf Egalitarian solution for the DPR utility model\footnote{In the context of the CREA project it was referred to as the ``Price-and-rate'' procedure}}
\begin{enumerate}[i.]
\item A mediator defines:
\begin{enumerate}[a.]
\item The number of levels in the ratings;
\item The market value for each good;
\item The rate of appreciation for each additional star in the rating (The parameter $K>1$ in the model).
\end{enumerate}
\item Each agent independently expresses a personal appreciation of each good at its market value using the defined rating scale.
\item The Egalitarian solution is computed.
\end{enumerate}

The Egalitarian allocation for the DPR utility model can be computed by means of the following linear programming problem.
\begin{equation}
\label{opt_prob}
\left\{
\begin{array}{rll}
\max & t &
\\
\mbox{s.t.} & \sum_{a \in A} K^{r_{ia}} m_a z_{ia} - w_i \left( \sum_{a \in A} K^{r_{ia}} m_a \right) t =  0 & \forall i \in N
\\
 & \sum_{i \in N} z_{ia} = 1 & \forall a \in G
 \\
 & z_{ia} \geq 0 & \forall i \in N, a \in G
\end{array}
\right.
\end{equation}
The formulation arises from the equivalence between the following two problems
\[
\max_{z \in \Phi(N,G)} \min_{ i \in N }\frac{\bar{U}_i(z)}{w_i} =
\max_{t, z} \left\{ t: \frac{\bar{U}_i(z)}{w_i} \geq t \: \forall i \in N, z \in \Phi(N,G)   \right\}
\]
and from a rescaling of the inequalities  in the second optimization problem. Moreover, I do not need to consider the median rating, because the utilities are normalized and, while 
 an inequality would be expected for each of the first $n$ constraints, each of them can be replaced by an equality, since \eqref{egaldef} holds in this context.
The feasible set is non-empty because the variables $z_{ia}=w_i$, $\forall a \in G, i \in N$ and $t=1$ satisfy all the constraints.
I now illustrate how the linear program is applied to one of the previous examples.
\begin{xmpl}[continues=xmpl1]
The linear program for this example becomes, after proper simplification
\begin{equation}
\label{opt_xmpl1}
\left\{
\begin{array}{rll}
\max & t &
\\
\mbox{s.t.} &  2 z_{11} + 2K z_{12} - \left( 1+K  \right)t=  0 & 
\\
 &  2 z_{21} + 2 K^3 z_{22} - \left( 1+K^{3}  \right)t=  0 & 
\\
 & z_{11}+z_{21} = 1 & 
 \\
 & z_{12}+z_{22} = 1 & 
 \\
 &z_{11},z_{12},z_{21},z_{22} \geq 0 & 
\end{array}
\right.
\end{equation}
It can easily be verified that the solution \eqref{xmpl:allocegal} and $t=2K^2/(2K^2-K+1)$ is feasible. The solution verifies the primal-dual complementarity conditions and is therefore optimal.

\end{xmpl}

Wilson \cite{wi98}, as reported by Sandomirskiy and Segal-Halevi \cite{ss19}, proved the existence of an egalitarian allocation of goods with $n-1$ sharings (i.e.\ split items). The result is unpublished\footnote{At the time of writing, Wilson \cite{wi98} cannot be retrieved via any popular search engine enquiry.}  and for ease of reference, I prove a statement that works for  the linear program defined in \eqref{opt_prob}.

\begin{prop}[Wilson \cite{wi98}]
\label{prop:basic}
An optimal basic solution for \eqref{opt_prob} identifies a solution with no more than $n-1$ split items.
\end{prop}
Remember that in a basic solution for a linear program, only a restricted group of basic variables equalling the number of constraints can be nonzero, and the simplex method moves from one basic solution to an adjacent one that differs from the first by only one variable, until the optimal allocation is reached.
\begin{proof}
Problem \eqref{opt_prob} has $ng+1$ variables and each basic solution has no more than $n+g$ nonzero variables. In the optimal solution, the variable $t$ is among the basic variables, because its value equates the nonzero utility to weight ratio level, and no more than $n+g-1$ variables in the optimal allocation $z^e$ can be greater than zero. Since each good $a \in G$ has to be assigned and, therefore, there exists $i \in N$ such that $z_{ia}$ is a basic variable, the number of split items cannot exceed the number of basic variables on $z$ minus the number of goods and there cannot be more than $n-1$ split items in the optimal solution.
\end{proof}

In principle, equalizing the agents' utilities may not seem the wisest choice because it requires a comparison of interpersonal utilities, which is a highly debated and questioned principle. In the present situation, however, agents' utilities are magnifications or contractions of the goods' market value and an allocation with equal (normalized) agents' utilities yields bundles of approximately equal market values for everyone, exact equality being problematic. A more detailed analysis reveals that differences in the bundles' market values can be explained in terms of the agents' satisfaction relative to the fraction of monetary value received. Some simple definitions are needed in order to provide a formal statement of this fact.

\begin{definition}
Given a feasible allocation $z$, the {\em monetary share} of agent $i$ is given by the ratio $\mu_i(z)/M$, while the {\em monetary share to entitlement (MSE)} ratio for the same agent is given by 
\[
\sigma_i(z)=\frac{\mu_i(z)}{w_i M}.
\]
\end{definition}
The MSE ratio compares the monetary share with the entitlement, and a value of 1 denotes an exact correspondence between the market value received and the entitlement in the division.
\begin{definition}
The {\em average utility per monetary share}, or in short, {\em the utility-per-money (UM) index} of the feasible allocation $z$ for agent $i$, $\nu_i(z)$, is defined as the ratio between the agent's normalized utility and the corresponding monetary share.
\begin{equation}
\label{ave_standutil}
\nu_i(z) = \frac{\bar{U}_i(z) M}{\mu_i(z)} .
\end{equation}
\end{definition}
The index indicates how much -- on average -- an amount of a good worth $1 \%$ of the total market value of goods contributes, as a percentage, to the normalized utility of an agent.
If an agent receives all goods, their UM is 1, and this is a benchmark for evaluating any other allocation: the higher the UM of an agent, the happier they are with each percentage of monetary share received.
Note that none of the indices introduced in the previous two definitions depend on the choice of currency used, since they are ratios of relative quantities. 

If $z^*$ is an Egalitarian allocation, then, for any two agents $i,h \in N$
\begin{equation}
\label{thm_mvave_three}
\nu_i(z^*) \sigma_i(z^*)=\nu_h(z^*) \sigma_h(z^*)
\end{equation}
The simple formula illustrates the inverse relationship between UM and MSE ratios: If $\nu_i(z^*) > \nu_h(z^*) $, then $\sigma_i(z^*)< \sigma_h(z^*)$ (and conversely). Most of all, it shows that the agents' UM indices and MSE ratios are listed in the opposite order in an Egalitarian allocation.

A monotone transformation of the UM index describes the quality of the allocated goods in terms of the agents' ratings, but some preliminary work is needed.
In order to make the agents' ratings comparable, they have to be centered around a reference value.
\begin{definition}
 The {\em central rating} for agent $i \in N$, $\bar{r}_i$, is defined as
\begin{equation}
\label{centrat_def}
\bar{r}_i = \log_K \left( \frac{\sum_{a \in A} K^{r_{ia}}  m_a}{M} \right) 
\end{equation}
The  rating difference from the central rating of a good $a \in A$ for agent $i \in N$ is defined by $r_{ia}- \bar{r}_i$.
\end{definition}
The difference will be a gain over the central rating, if positive -- or a loss, if negative.
Note that the central rating satisfies the following identities
\begin{equation}
\label{centrat_property}
\sum_{a \in A} K^{(r_{ia}-\bar{r}_i)}m_a =M  \qquad i \in N.
\end{equation}
Also note that the magnitude of $\bar{r}_i$ gives only an indication of the rating levels: an agent that assigns high ratings to valuable goods will have a higher central rating than that of one who assigns lower ratings or another one who assigns high ratings to less valuable items. The difference, however, will not mean that an agent has rights to a larger or smaller share in the division procedure.  

The following result gives an alternative characterization of  the UM index: it is the average of the personal magnifying factors for the goods, weighted with the monetary values of the portions of goods received by the agent.
\begin{prop}
\label{aveUM}
For any feasible $z$, we have
\begin{equation}
\label{ave_standutil}
\nu_i(z) = \frac{\sum_{a \in A} K^{(r_{ia} - \bar{r}_i)} m_a z_{ia}}{\sum_{a \in A}  m_a z_{ia}} \qquad i \in N .
\end{equation}
\end{prop}
\begin{proof}
Considering a rating scale with $2q+1$ steps, we have
\begin{multline*}
\nu_i(z)= \frac{M \sum_{a \in A}K^{(r_{ia} - q-1)} m_a z_{ia}}{\mu_i(z) \sum_{a \in A}K^{(r_{ia} - q-1)} m_a} =
\\
 \frac{M \sum_{a \in A}K^{(r_{ia} - \bar{r}_i)} m_a z_{ia}}{\mu_i(z) \sum_{a \in A}K^{(r_{ia} - \bar{r}_i)} m_a} =  \frac{\sum_{a \in A}K^{(r_{ia} - \bar{r}_i)} m_a z_{ia}}{\sum_{a \in A}  m_a z_{ia}} 
\end{multline*}
For the first equality, I apply the definition of $\nu_i$, for the second, I multiply and divide by $K^{q+1-\bar{r}_i}$ and for the third, I apply \eqref{centrat_property} and the definition of $\mu_i(z)$.  
\end{proof}
The previous result maks it possible to characterize the quality of the bundles in terms of ratings.
\begin{definition}
The {\em average difference from the central rating per monetary share}, or, in short, the rating difference (RD) index of the feasible allocation $z$ for agent $i$ is defined as:
\begin{equation}
\label{ave_rate}
\rho_i(z) = \log_K(\nu_i(z)) 
\end{equation}
\end{definition}
Following Proposition \ref{aveUM}, the RD index is the average of the goods' standardized rating weighted with the monetary values of the portions of goods received by the agent.
Recall that the RD index for the whole set of goods is 0, and, consequently, also a random allocation of goods (and fractions thereof) worth a fixed amount of market value yields a null average.  The index $\rho_i(z)$ will therefore indicate the average difference between the agent's rating obtained from the allocation $z$ and that obtained from a random bundle of the same value.

Since the RD index is a monotone transformation of the UM index, both measures rank the players in the same order. 
When all the agents have equal entitlements, monetary values and MSE ratios also agree in the ranks; however, given equation \eqref{thm_mvave_three}, the two pairs of indices rank agents in reverse order.
I now apply the indices just defined in the previous examples.

\begin{xmpl}[continues=xmpl1]
For any $K>1$, the Egalitarian solution is defined by \eqref{xmpl:allocegal} with the common utility obtained by the agents given by \eqref{xmpl:utilegal}. Now,
$\bar{r}_1=\bar{r}_2=3$ and the indices that measure the quality of the Egalitarian solution for each agent are described in Table \ref{table:xmpl1}, where the general formula and the numerical values for $K=1.2$.
\begin{table}[h!]
\caption{Characterization of the solution for Example \ref{xmpl1}.}
\label{table:xmpl1}
\begin{center}
\subfloat[general formula]{
 \begin{tabular}{||c||c| c||} 
 \hline
 Agent & $\mu_i(z^e)$ & $\rho_i(z^e)$  \\
 \hline\hline
 $I$ & $\frac{3 K^3 -2 K^2 +2 K-1}{2 K^3 - K^2 +K}$ &   $\log_K \left(\frac{2K^3}{3K^3-2K^2+2K-1}  \right)$
\\ 
 \hline
$II$ & $\frac{K^3+1}{2K^3-K^2+K}$ &  $\log_K \left( \frac{2K^3}{K^3+1}  \right)$
\\ 
 \hline
\end{tabular}}
\subfloat[$K=1.2$]{
 \begin{tabular}{||c||c| c||} 
 \hline
 Agent & $\mu_i(z^e)$ & $\rho_i(z^e)$  \\
 \hline\hline
 $I$ & $1.152$ &   $-0.3801$
\\ 
 \hline
$II$ & $0.848$ &  $1.2974$
\\ 
 \hline
\end{tabular}}
\end{center}
\end{table}
As expected, the differences in monetary values are compensated by the difference in perceived quality from the received items. 
When $K=1.2$, the smaller monetary share received by agent $II$ is compensated by a larger RD index. By receiving the least valuable item, the agent will get an average increase of 1.2974 stars over any random allocation worth $0.848$. Conversely, agent $I$ will receive a bundle worth 1.152 that, when compared with a random bundle of the same value is worth 0.3801 stars less. 
\end{xmpl}

\begin{xmpl}[continues=xmpl:toy_Moulin]
If the UM and RD indices are plotted for the four agents, as $K$ ranges in $[1,1.5]$, Figure \ref{fig:avut} is obtained.
\begin{figure}
     \centering
     \begin{subfigure}[b]{0.45\textwidth}
         \centering
         \includegraphics[width=\textwidth]{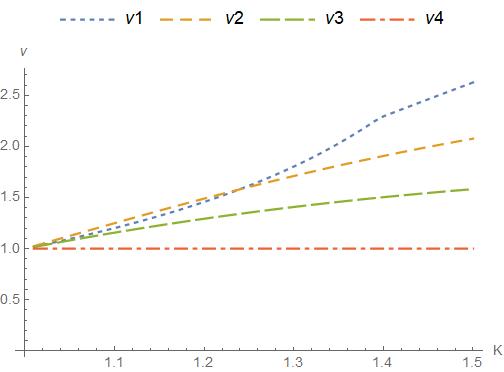}
         \caption{The UM index.}
         \label{fig:avut}
     \end{subfigure}
     \hfill
     \begin{subfigure}[b]{0.45\textwidth}
         \centering
         \includegraphics[width=\textwidth]{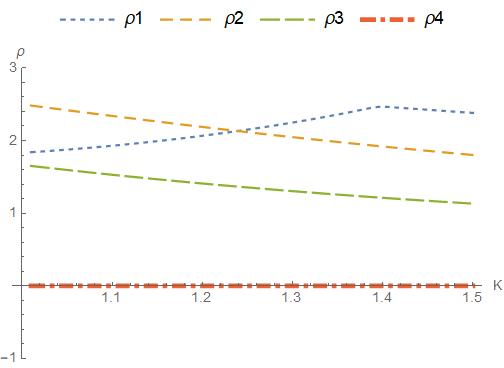}
         \caption{The RD index.}
         \label{fig:rho}
     \end{subfigure}
        \caption{Goods' appreciation in Example \ref{xmpl:toy_Moulin}.}
        \label{fig:aveAS_Moulin}
\end{figure}
As expected, the curves' rank in both graphs is identical and it is opposite to that of the monetary values -- and the lines of agents $I$ and $II$ cross at $K \approx 1.2391$. 
\end{xmpl}

A positive RD index denotes a gain over the central rating. Example \ref{xmpl1}, however, shows that an RD  index can be negative -- denoting a loss over the central rating.

From a global perspective, however, the loss cannot affect all the agents.

\begin{prop}
\label{avenu}
The average $\bar{\nu}(z^e)$ of the agents' UM indices, weighted with their bundles' monetary values is always greater then or equal to 1.
\end{prop}
\begin{proof}	
The Egalitarian solution $z^e$ has the Fair Share Guarantee. This and the definition of the UM index yield
\begin{equation}
\label{eq:avenu1}
\frac{\nu_i(z^e) \mu_i(z^e)}{M} = \bar{U}_i(z^e) \geq w_i \qquad \mbox{for every } i \in N
\end{equation}
Now, according to \eqref{eq:avenu1} and to the fact that $\sum_{i \in N} \mu_i(z^e)=M$, I can write
\[
\bar{\nu}(z^e) = \frac{\sum_{i \in N} \nu_i(z^e) \mu_i(z^e)}{\sum_{i \in N} \mu_i(z^e)} = \frac{\sum_{i \in N} \bar{U}_i(z^e) M}{M} \geq \sum_{i \in N} w_i =1 .
\]
\end{proof}
In the same spirit of the previous definitions, $\bar{\rho}(z)=\log_K(\bar{\nu}(z))$ can be considered a valid centrality index for the RD indices. By Proposition \ref{avenu}, $\bar{\rho}(z^e)$ is always non-negative.

\begin{xmpl}[continues=xmpl1]
It can be easily shown that 
\[
\bar{\nu}(z^e)=\frac{2K^2}{2K^2-K+1} >1.
\]
Therefore $\bar{\rho}(z^e)>0$.
\end{xmpl}

\subsection{Assessing a value for $K$}
The implemented procedure relies on determining a value for the parameter $K$. How should it be chosen? First of all, I examine what happens when $K$ approaches the extreme values of its range. As $K$ gets arbitrarily large, each agent's utility narrows down to goods with the  highest rating. Let $G^{\max}_i$ be the set of goods that received top ratings from agent $i$. It is easy to prove the following:

\begin{prop}
When $w_i=n^{-1}$ for every $i \in N$, as $K \to \infty$, the normalized utility of good $a$ for agent $i$ converges to
\begin{equation}
\label{unorm_infty}
\bar{u}_{ia} \to 
\begin{cases}
\cfrac{m_a}{\sum_{b \in G_i^{\max}} m_{bi}} & \mbox{if } a \in G^{\max}_i
\\
0 & \mbox{otherwise}
\end{cases}
\end{equation}
\end{prop}
Considering the asymptotic values of the agents' optimal utilities, I focus on a closed formula for the special case where agents have equal weights and there is a common set of goods, $G^{\max}_c$, which are top-rated by all the agents, and if two agents gave their top rating to a good, then all the agents did, i.e.\ $G^{\max}_i \cap  G^{\max}_j = G^{\max}_c$ for every $i,j \in N$. Let $z^e(K)$ be an Egalitarian solution associated with a given value $K$.

\begin{prop}
When $w_i=n^{-1}$ for every $i \in N$, as $K \to \infty$,
\begin{equation}
\label{form:unorm_asint}
\bar{U}_i(z^e(K)) \to 1- \cfrac{(n-1)\prod_{i \in N} d_i}{\sum_{i \in N} \prod_{j \neq i} d_j}\quad \mbox{where }d_i=\cfrac{\sum_{a \in G^{\max}_c} m_a}{\sum_{a \in G^{\max}_i} m_a} \: \forall i \in N
\end{equation}
\end{prop}
\begin{proof}
As $K \to \infty$, the normalized utility of good $a$ for agent  $i$ converges as indicated in \eqref{unorm_infty}. As $K$ grows, in order to have an efficient allocation, with the normalized total utility of all the agents equal, goods in $G^{\max}_i \setminus G^{\max}_c$ must be assigned to agent $i$, goods outside $G^{\max}_i$ can be assigned to any agent since their contribution in utility becomes negligible, while goods in $G^{\max}_c$ must be distributed (and possibly split) among all the agents in order to keep the total utilities equal. As $K \to \infty$, the normalized utility for agent $i$ of the goods converges to $d_i$ on $G^{\max}_c$, and to $1 - d_i$ on $G^{\max}_i \setminus G^{\max}_c$. To keep equality among the utilities, the fraction $t_i$ of $G^{\max}_c$ that is given to each agent $i$ is computed as the solution of the following linear system
\begin{equation}
\label{ti:sys}
\begin{cases}
t_1 d_1 - t_j d_j = d_1 - d_j & j=2,\ldots,n
\\
\sum_{i \in N} t_i=1 &
\end{cases}.
\end{equation}
It turns out that
\begin{equation}
\label{ti:sol}
t_i =\frac{\sum_{j \neq i} \prod_{h \neq j} d_h - (n-2) \prod_{h \neq i} d_h}{\sum_{j \in N} \prod_{h \neq j} d_h} \qquad \mbox{for every } i \in N.
\end{equation}
To show that \eqref{ti:sol} is the solution of \eqref{ti:sys}, we rewrite the first $n-1$ equations in the system as $d_1(1-t_1)=d_j(1-t_j)$. Now 
\[
d_i(1-t_i)=\frac{(n-1) \prod_{h \in N} d_h}{\sum_{j \in N} \prod_{h \neq j} d_h} \qquad \mbox{for every } i \in N.
\]
To verify the last equation in \eqref{ti:sys}, the sum of all the denumerators in \eqref{ti:sol} gives the denominator, and therefore $\sum_{i \in N}t_i=1$. The common utility level $1-d_i+ t_i d_i$ yields \eqref{form:unorm_asint}.
\end{proof}

When there is no overlap in the top ratings, i.e.\ $G^{\max}_c = \varnothing$, then $\bar{U}_i(z^*) \to 1$ as $K \to \infty$. Furthermore, when $n=2$, then \eqref{form:unorm_asint} becomes
\[
\bar{U}_i(z^e(K)) \to 1 - \cfrac{d_1 d_2}{d_1+d_2} \qquad \mbox{as } K \to \infty.
\]
For all the other cases, i.e.\ when some goods received a top rating from some (but not all) the agents, or when agents have different weights, the asymptotic value can be computed by means of a system of linear equations, but providing a closed form formula is too complicated and probably useless for the number of different cases that should be considered.

When $K$ is large, only the top rated items count, and items with lower ratings are unable to compensate for gaps in the required utility level. As $K$ grows, a paradox may occur: items can be assigned to agents with infinitesimal utility. This is particularly apparent in the case of unequal entitlements.

\begin{xmpl}
\label{xmpl:paradox}
Consider agents $I$ and $II$ with weights $w_I=6/7$, $w_{II}=1/7$ and 3 goods ($A$, $B$, $C$). The goods' monetary values and ratings are given in Table \ref{table:paradox}. 
\begin{table}[h!]
\begin{center}
\caption{The ratings for Example \ref{xmpl:paradox}}
\label{table:paradox}
\begin{tabular}{||c|c||c|c|c||}
\hline \hline
Item & Market v.\ & $I$ & $II$  \\
\hline \hline
$A$ & 100 & \ppstab  & \pstab  \\
\hline
$B$ & 800 & \nstab  & \mmstab  \\
\hline
$C$ & 100 & \mmstab  & \ppstab \\
\hline
\end{tabular}
\end{center}
\end{table}

When $1 < K \leq K_0 \approx 1.14727$, agent $I$ gets item $B$, agent $II$ gets item $C$ and the two agents share item $A$, with the share of the first agent increasing to 1. When $K > K_0$, agent $I$ gets items $A$ and $B$, while the two agents share item $C$, with the share of agent $I$ converging to 5/6 as $K \to \infty$. While agent $I$ gets most of item $C$ as $K$ grows, the utility of this item for the agent becomes negligible, as it becomes $O\left(K^{-4}\right)$ as $K \to \infty$. With a value as low as $K=2$, for instance, agent $I$ receives a share of item $C$ approximately equal to $0.66892$, ashare that will count for an approximate amount of $0.00607$ of the normalized utility of the agent. 
\end{xmpl}
The above example shows an instance where setting $K$ even less than one unit away from 1 may not be a good idea. To find out what happens at the other end of $K$'s range, the following continuity result is helpful.

\begin{prop}
\label{cont_linprog}
Suppose $\{K_h\}$ is a converging sequence of nonnegative real numbers $K_h \to K$ as $h \to \infty$, then, for every $i \in N$
\begin{equation}
\label{cont_barU}
\bar{U}_i(z^e(K_h)) \to \bar{U}_i(z^e(K)) \qquad \mbox{as }h \to \infty
\end{equation}
\end{prop}
\begin{proof}
It is easy to show that both the linear programming problem  \eqref{opt_prob} written in canonical form (i.e., with all the constraints expressed as inequalities together with the nonnegativity of all the variables), and its dual, satisfy the hypotheses of Proposition 8 in Wets \cite{we85}. Then, Theorem 2 from the same reference holds, and the optimal value of both the primal and the dual is continuous with respect to the parameter $K$.
\end{proof}

If $K=1$, every piece of information about the agents' preferences is lost, and any division of the goods in which the bundles' monetary value is proportional to the weights is optimal.   When $K$ approaches 1 from the right, the agents' normalized:utilities for the Egalitarian solution converge to the corresponding entitlements. 
\begin{prop}
As $K \to 1^+$, $U_i(z^e(K)) \to M w_i$ and $\bar{U}_i(z^e(K)) \to w_i$ for every $i \in N$.
\end{prop}
\begin{proof}
Apply Proposition \ref{cont_linprog} and  note that $K \to 1^+$, the IPS shrinks to a line, and the utility of every good coincides with its market value. 
\end{proof}
Consider a sequence $\{K_h\}$ with $K_h \to 1^+$ as $h \to \infty$. If  $z^e(K_h) \to z^*$ as $h \to \infty$, then, according to Proposition \ref{cont_linprog}, the limit Egalitarian solution has all the goods with the same utility for every agent (given by their monetary value). If the allocation remains the same for $h$ large, the limit solution keeps the same structure. 
\begin{xmpl}[continues=xmpl:paradox]
As $K \to 1^+$, the Egalitarian solution converges to 
\[
z^e(K) \to z^{e}=
\kbordermatrix{
     & A & B & C
     \\
     $I$ & 4/7 & 1 & 0  
     \\
     $II$ & 3/7 & 0 & 1
},
\]
with $\mu_1(z^{e})=6000/7$ and $\mu_2(z^{e})=1000/7$. The allocation scheme is the same as that for $1 < K \leq K_0 \approx 1.14727$ and only the split proportions change. 
\end{xmpl}

The solution looks reasonable for the previous example but may not distinguish between goods with high or low ratings as long as the difference in the agents' ratings remains the same.

\begin{rema}
In all the examined examples, $\bar{U}_i(z^e(K))$ turns out to be strictly increasing in $K$ for every $i \in N$. Whether this is always true remains an open question.
\end{rema}

From the previous discussion and examples, it emerges that $K$ should be chosen close enough to 1, so that even goods with lowest rating retain enough value, but not too close or exactly equal to 1, so that differences in the ratings count. I propose a simple elicitation procedure in which agents are asked to evaluate the interval of admissible utility values and compute the ratio between the interval's end points. 
Agents may evaluate this ratio jointly or separately and, in the latter case, the average value is computed. Denoting  such a ratio as $R$ and considering a rating scale with $2q+1$ steps, it would be natural to set $K=\sqrt[2q]{R} $. For the applications that follow, I will use a 5-star rating system, thus $q=2$. I also assume that a top-rated good is worth 50\% more than a least-rated one. Therefore, I will assume $K=\sqrt[4]{3/2} \approx 1.10668$.

In the proposed procedure, agents rate each good individually, with no constraint on the total number of stars that each agent can assign. The following example illustrates why putting bounds on the total number of stars may be a bad idea.
\begin{xmpl}
\label{xmpl_constrating}
Consider an instance with 4 agents ($I$, $II$, $III$ and $IV$) and 4 goods ($A$, $B$, $C$ ad $D$) -- each with  a price tag of 100. Suppose that each of the first three agents is very interested in one good: agent $I$ in good $A$, agent $II$ in good $B$, agent $III$ in good $C$, and they all have a mild interest in good $D$. Finally, suppose that agent $IV$ is interested in goods $A$, $B$ and $C$, but has very little interest in good $D$.  If agents were able to assign the ratings freely, the preferences shown in Table \ref{table_unconst_ratings} are what we would expect. 
\begin{table}
\caption{The data for Example  \ref{xmpl_constrating}}
\begin{subtable}[h]{0.75\textwidth}
\centering
\begin{tabular}{||c|c||c|c|c|c||}
\hline \hline
Item & Market v.\ & $I$ & $II$ & $III$ & $IV$ \\
\hline \hline
$A$ & 100 & \ppstab  & \mmstab & \mmstab & \ppstab  \\
\hline
$B$ & 100 & \mmstab  & \ppstab  & \mmstab  & \ppstab\\
\hline
 $C$ & 100 & \mmstab  & \mmstab & \ppstab  & \ppstab \\
\hline
 $D$ & 100 & \mstab  & \mstab & \mstab  & \mmstab \\
\hline
\end{tabular}
\caption{Unconstrained ratings}
\label{table_unconst_ratings}
\end{subtable}
\hfill
\begin{subtable}[h]{0.2\textwidth}
\centering
\begin{tabular}{||c||}
\hline \hline
 $IV$ \\
\hline \hline
 \nstab  \\
\hline
 \nstab\\
\hline
 \nstab \\
\hline
 \mmstab \\
\hline
\end{tabular}
\caption{Constrained}
\label{table_const_ratings}
\end{subtable}
\end{table}

If, instead, agents were only allowed to distribute 10 stars among the goods, the profile for agent $IV$ would change into the one described in Table \ref{table_const_ratings}.
Let $z^u$ be the Egalitarian allocation resulting from the market value and the rating profiles in Table \ref{table_unconst_ratings}, together with $K=\sqrt[4]{3/2}$, and let $z^r$ be the Egalitarian allocation, in which the ratings profile for agent $IV$ has been modified as in Table \ref{table_const_ratings}, due to the constraint. We get
\begin{gather}
\label{unconstr_alloc}
z^u=
\kbordermatrix{
     & A & B & C & D 
     \\
     $I$ & 0.8912 & 0 & 0 & 0  
     \\
     $II$ & 0 & 0.8912 & 0 & 0 
      \\
     $III$ & 0 & 0 & 0.1535 & 1
          \\
     $IV$ & 0.1088 & 0.1088 & 0.8465 & 0
}
\\
\notag
z^r=
\kbordermatrix{
     & \phantom{A} & \phantom{B} & \phantom{C} & \phantom{D} 
     \\
     $I$ & 0.8995 & 0 & 0 & 0  
     \\
     $II$ & 0 & 0.8995 & 0 & 0 
      \\
     $III$ & 0 & 0 & 0.8995 & 0
          \\
     $IV$ & 0.1005 & 0.1005 & 0.1005 & 1
}
\end{gather}
Based on the data symmetry, other solutions for the unconstrained problem are obtained by assigning good $D$ to agents $I$ or $II$, with the fractions described in \eqref{unconstr_alloc}.
It would be hard to describe the outcome $z^r$ as satisfactory for agent $IV$. Note that if Table \ref{table_const_ratings} were the true preference profile for agent $IV$, solution $z^c$ would be justified by the fact that the difference  between the ratings of good $D$ and the other three goods is moderate -- and would explain the assignment of the least favoured item in return for a larger market value.

\end{xmpl}

\subsection{Model Robustness}
Every model that aims at describing the utility of a group of agents using a common interpretative key relies on assumptions that allow for an elicitation process that is bearable for the agents. 

The inevitable cost of this construction is a simplification and a homogeneity that may cut out some details of the rational process.A reality check is particularly important for the DPR utility model introduced here since it relies on facilitating the revelation of preferences by the agents, with the aim of addressing the largest possible population of potential users and making them aware of the procedure’s functioning.

A validation of the model's credibility must rely on a proper experimental setting. A first encounter with the economics labs can be found in \cite{ddm19}. The experiment refers to an early stage of the project, in which students at Luiss University expressed their preferences for a set of goods and were asked to choose between the Egalitarian and the Nash/Competitive solutions -- or to reject the solution altogether and ``go to court'' (i.e.\ receive a curtailed payoff).\footnote{For the record, the Nash/Competitive solution won by $2\%$. A margin this narrow does not warrant the title of a statistically certified winner.}

A resumption of the laboratory work to test the validity of the DPR model is out of the scope of the present work. Instead, I rely on simulations to test the parameters' robustness. I therefore assume the DPR model to be valid, but with a misspecification of the chosen parameters. I assume the goods' market values to be assessed by an impartial expert -- or to be agreed upon by the agents themselves. In both cases the agents should not or could not complain about the outcome. The focus is on two other issues:
\begin{enumerate}[1.]
\item How wrong is the model when too few rating levels are specified?
\item How wrong is the model when the magnification factor $K$ is different from the ``true'' one describing  the agents' utility?
\end{enumerate}
Regarding the first question, I examine what happens when the agents' utility cannot be captured by the usual 5-levels rating scale, because the agents are capable of calibrating their preferences over a richer discrete range. I consider a larger rating scale with 11 levels. The choice is motivated by the fact that, if the magnifying factors are chosen to guarantee the same utility for the highest and lowest ranking in both rating scales, then only 3 levels of the largest rating scale have a perfect equivalent in the smallest scale. These are levels 1, 6 and 11 in the larger scale which correspond to  levels 1, 3 and 5 in the smaller one. The other 8 levels of the larger scale do not have  a perfect matching in the smaller scale. Nevertheless, a discerning and rational agent who is asked to use the smaller scale, will proceed by proximity, grouping the richer rating scale as $\{\{1,2\},\{3,4\},\{5,6,7\},\{8,9\},\{10,11\}\}$.

\begin{figure}[h!]
\begin{center}
\caption{The distribution of relative errors in coarsening the rating scale.}
\label{error11to5:hist}
\includegraphics[width=0.8\textwidth]{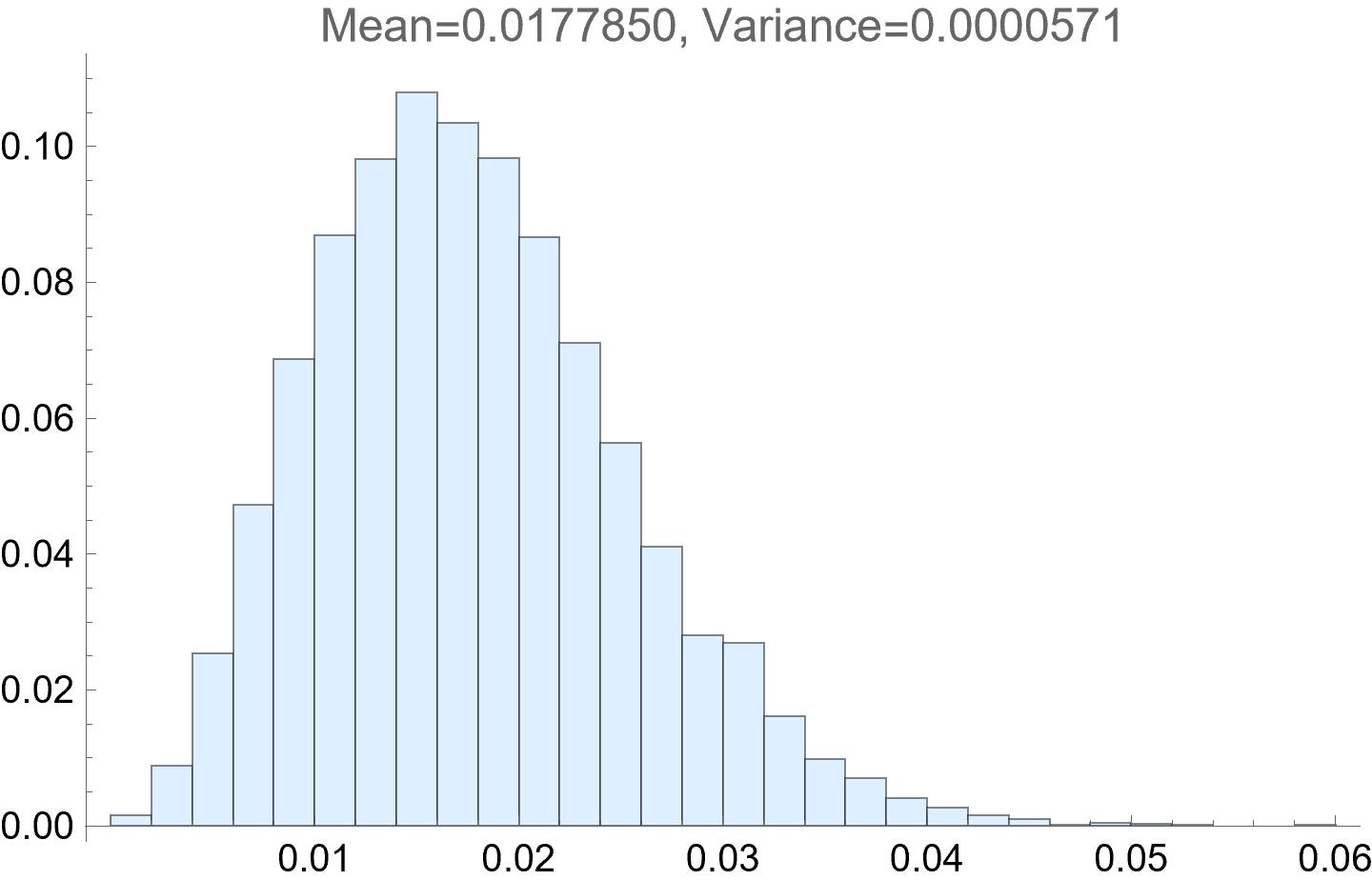}
\end{center}
\end{figure}

Using Mathematica 12.3, I  ran a simulation where the following situation is repeated $10,000$ times: 8 items, each with a random market value ranging between 100 and $1,000$ (euros) are contended among 4 equally important agents, whose ratings are randomly assigned on the 11-level rating scale. Each utility is transformed into the 5-level rating scale by proximity, but is evaluated according to the original larger scale. The normalized utilities of the 4 agents are no longer equal to each other, and the allocation is no longer optimal because one or more agents show a utility level lower than the Egalitarian value computed for the larger rating scale. The gap between the utility of the Egalitarian solution for the true model and the lowest of the agents' utilities (i.e.\ the value of the suboptimal allocation according to the objective \eqref{egsol}) is recorded. To facilitate the comparison of errors among the simulations, the gap is divided by the Egalitarian level to return a relative loss.  Figure \ref{error11to5:hist} shows a histogram of the distribution loss in the simulations. The data shows an average loss of $2 \%$ on the true optimal value.

I turn now to the misspecification of the magnification factor $K$. Again, 8 items, each with a randomly assigned market value between 100 and $1,000$ (euros) are divided among 4 equally important agents whose randomly assigned ratings are correctly described by a  5-level rating scale. The announced value is $K=\sqrt[4]{3/2}$, but the true value differs by a random gap. Allocations are computed using the announced value, but evaluated by the true value of $K$. Again, the computed allocation is no longer optimal and agents have different true normalized utilities, and the relative error is computed as the difference between the (true) Egalitarian value and the lowest reported utility divided by the  Egalitarian value.

\begin{figure}
     \centering
     \begin{subfigure}[b]{0.48\textwidth}
         \centering
         \includegraphics[width=\textwidth]{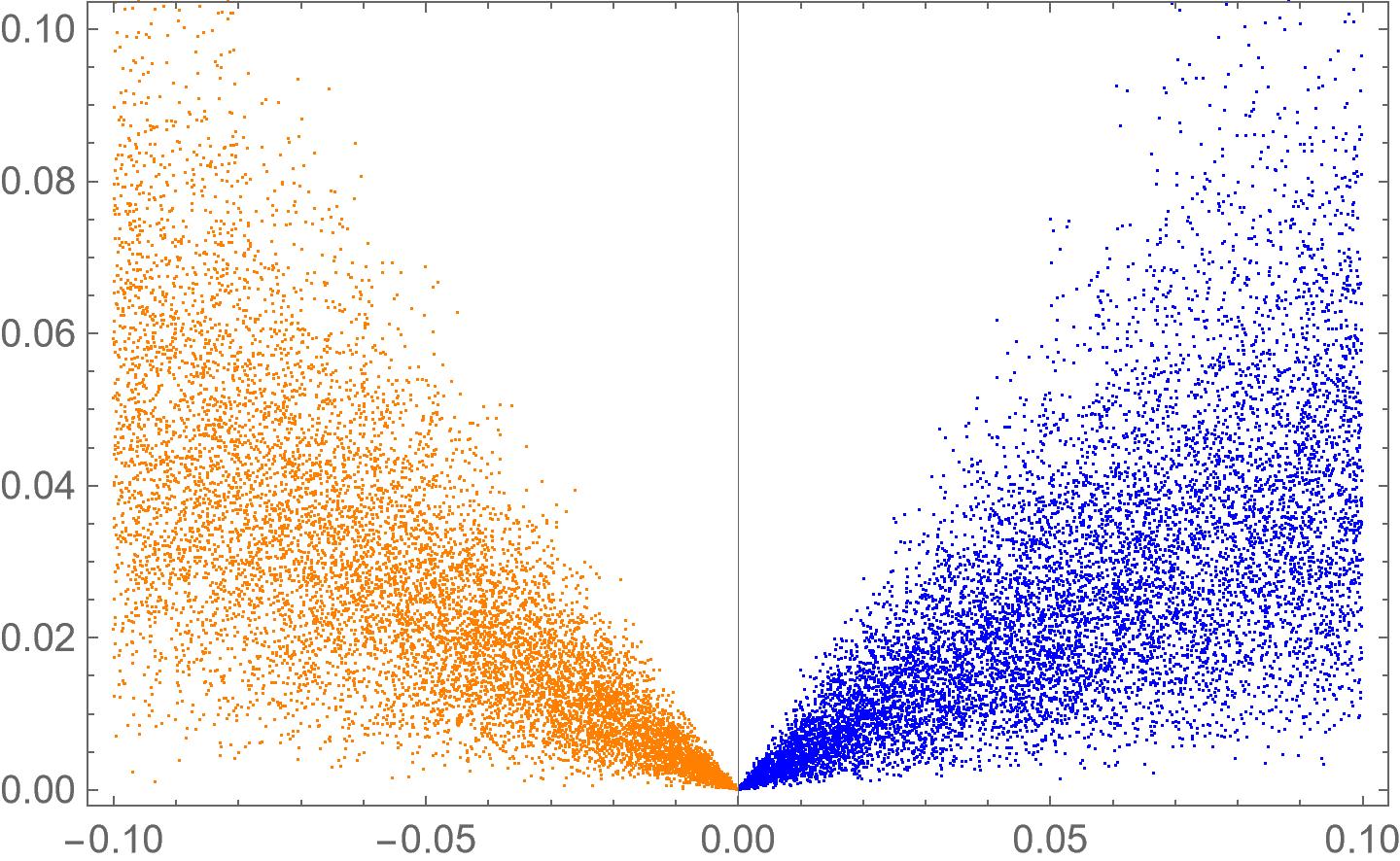}
         \caption{Relative errors ($y$) vs.\ deviation ($x$)}
         \label{fig:posnegplot}
     \end{subfigure}
     \hfill
     \begin{subfigure}[b]{0.48\textwidth}
         \centering
         \includegraphics[width=\textwidth]{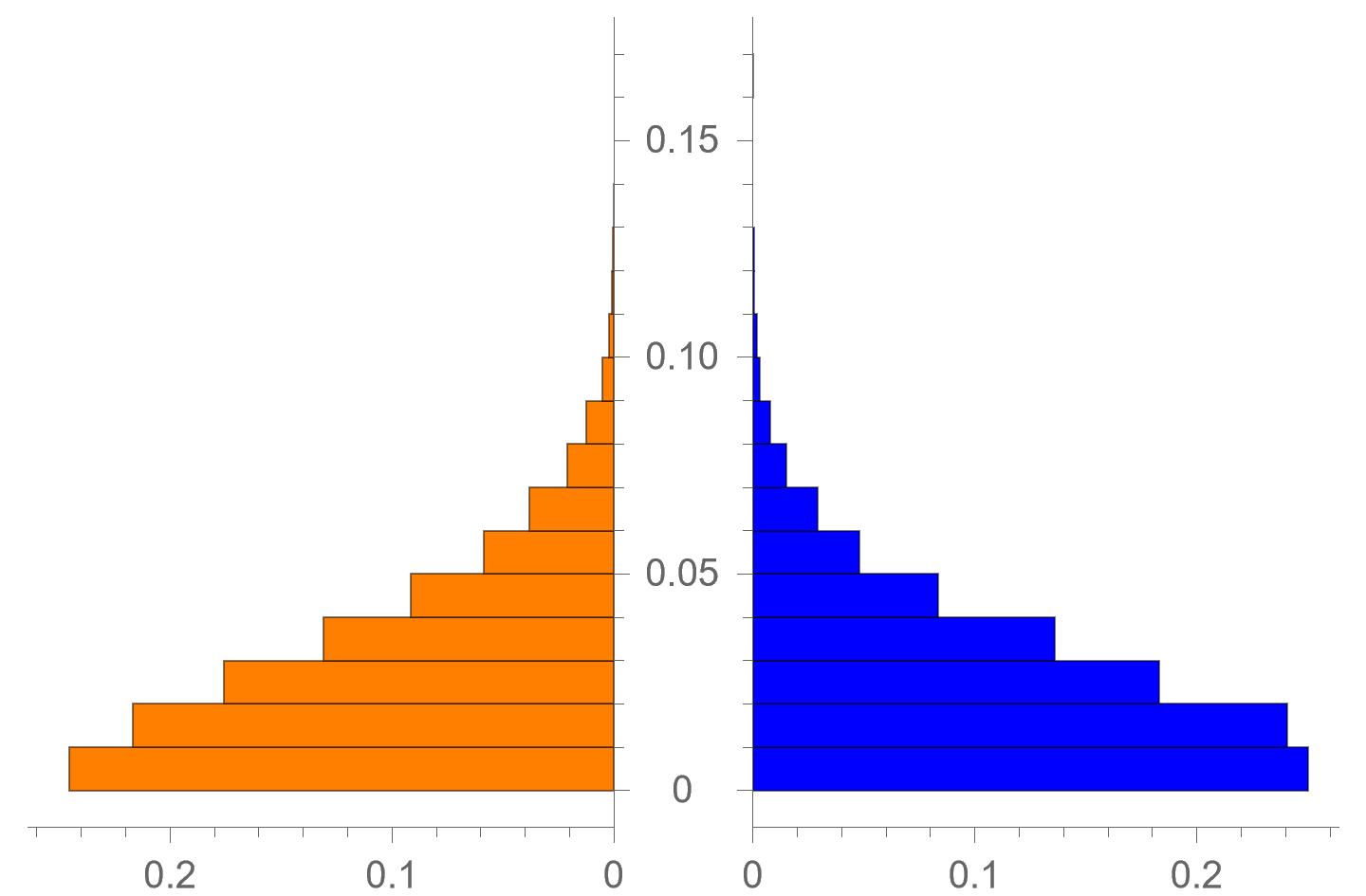}
         \caption{Distribution of errors:  positive (blue) vs.\ negative (orange) deviations}
         \label{fig:posnegcomp}
     \end{subfigure}
        \caption{Simulation results with a small two-sided gap between the true and the announced value of $K$.}
        \label{fig:posneg}
\end{figure}

A first batch of $20,000$ random instances with a gap in the range $[-0.1,0.1]$ was launched and the results are shown in Figure \ref{fig:posneg}. Figure \ref{fig:posnegplot} plots the relative errors against the magnitude and the direction of the deviation. Figure \ref{fig:posnegcomp} compares the distribution of errors from positive deviations with those from negative ones. The Kolmogorov-Smirnov Test rejects the hypotheses that the two sets of data are generated by the same probability distribution. However, the descriptive statistics are very similar: the distribution of the errors from positive deviations has a mean of $0.024813$ and a variance of $0.0003706$, while the distribution from negative deviations has a mean of $0.026702$ and a variance of $0.0004393$.  

\begin{figure}
     \centering
     \begin{subfigure}[b]{0.48\textwidth}
         \centering
         \includegraphics[width=\textwidth]{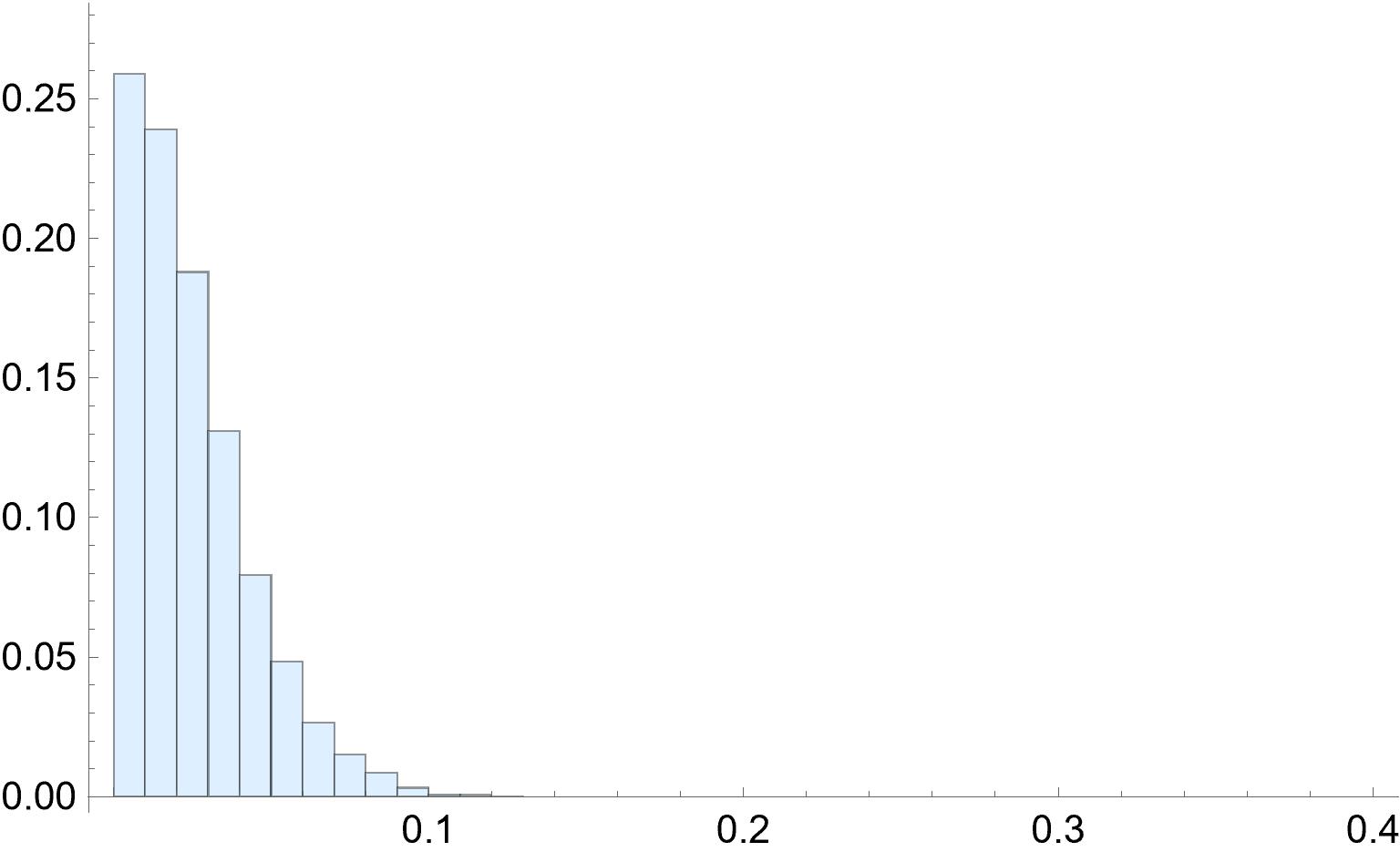}
         \caption{Deviation in $[0,0.1]$}
         \label{fig:histclass1}
     \end{subfigure}
     \hfill
     \begin{subfigure}[b]{0.48\textwidth}
         \centering
         \includegraphics[width=\textwidth]{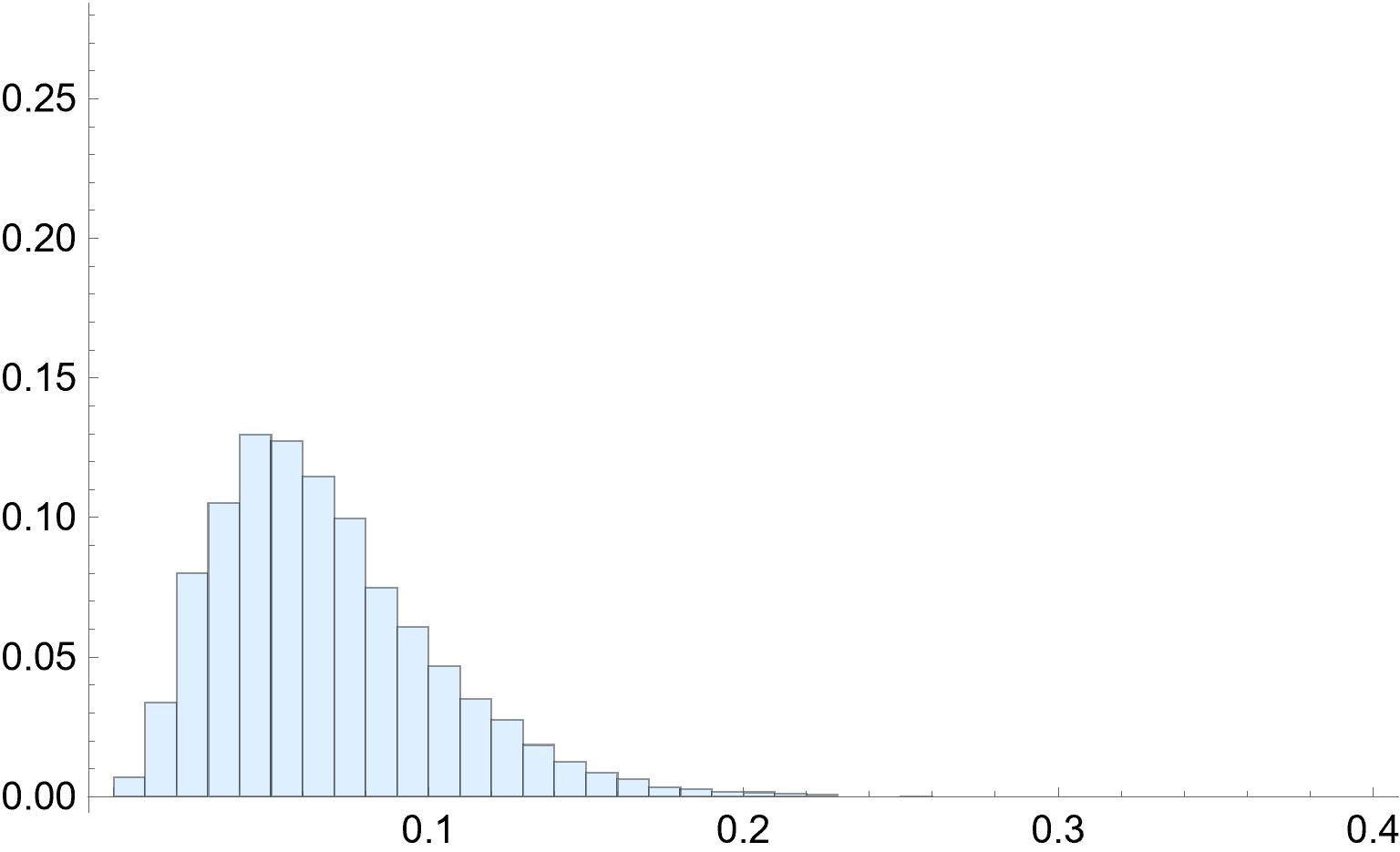}
         \caption{Deviation in $[0.1,0.2]$}
         \label{fig:histclass2}
     \end{subfigure}     \\
     \begin{subfigure}[b]{0.48\textwidth}
         \centering
         \includegraphics[width=\textwidth]{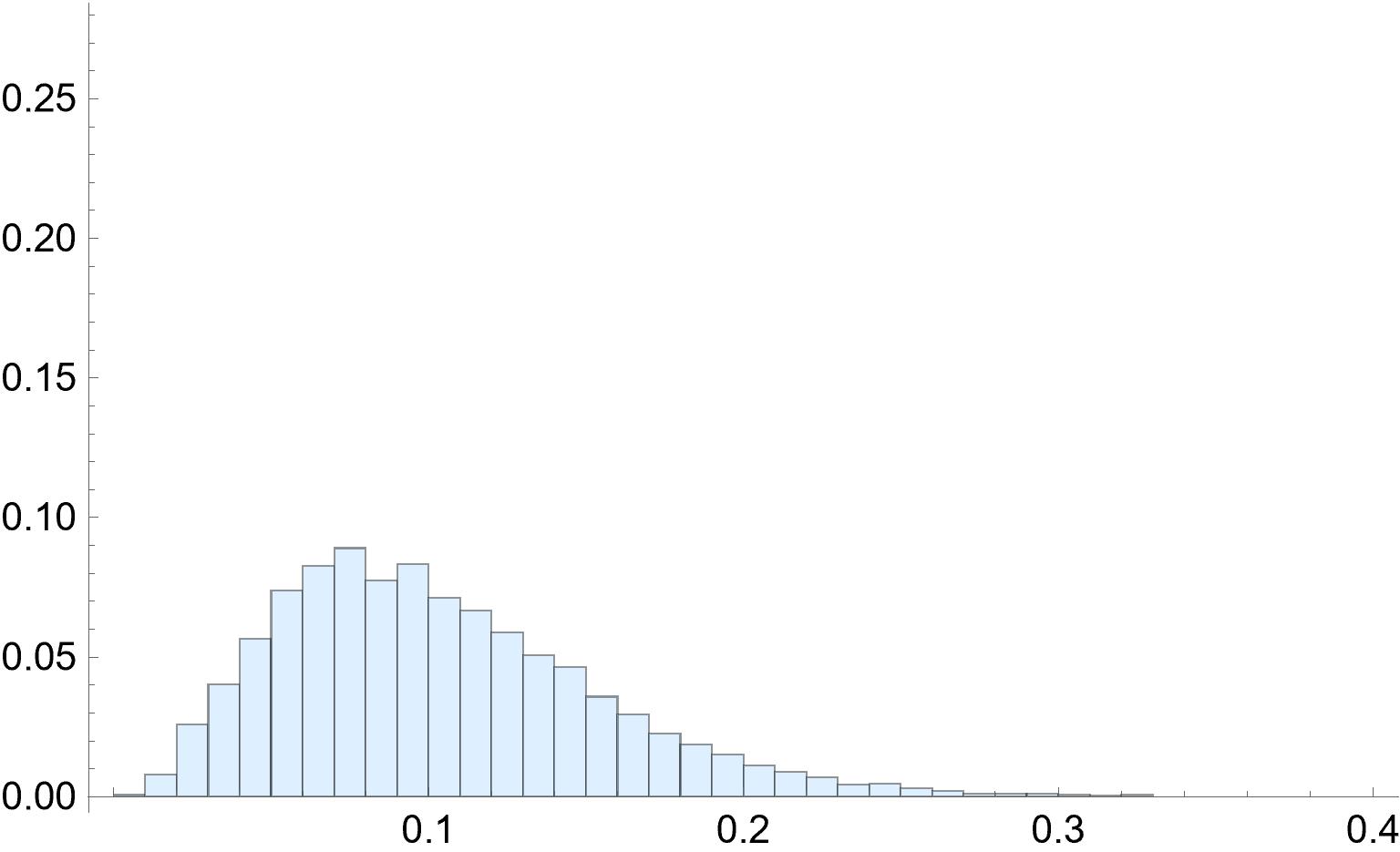}
         \caption{Deviation in $[0.2,0.3]$}
         \label{fig:histclass3}
     \end{subfigure}
     \hfill
     \begin{subfigure}[b]{0.48\textwidth}
         \centering
         \includegraphics[width=\textwidth]{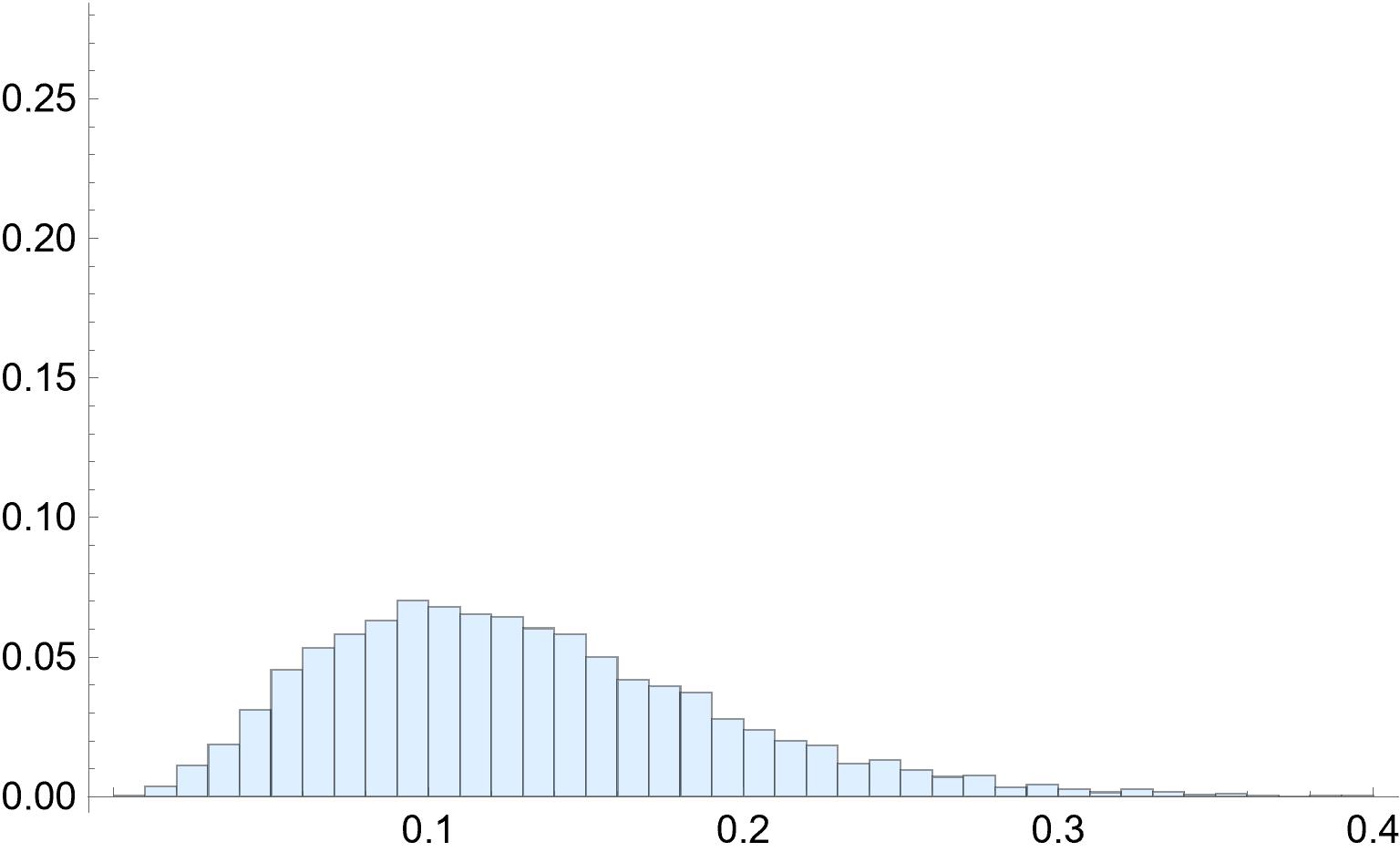}
         \caption{Deviation in $[0.3,0.4]$}
         \label{fig:histclass4}
     \end{subfigure}      
        \caption{Simulation results with an increasing gap between the true and the announced value of $K$.}
        \label{fig:histclass}
\end{figure}

A second batch of $20,000$ simulations considers a positive random deviation not greater than $0.4$. The data is then divided into 4 groups according to the magnitude of the deviations. The histograms of the four distributions are shown in Figure \ref{fig:histclass}. The descriptive statistics for the four distributions are shown in Table \ref{table:histclass}. Both the mean and the variance grow with the gap, with a linear growth for the mean. 

\begin{table}[h!]
\begin{center}
\caption{Descripive statistics for the distributions in Figure \ref{fig:histclass}}
\label{table:histclass}
\begin{tabular}{||c||c|c||}
\hline \hline
Dev.\ Magnitude & Mean & Variance  \\
\hline \hline
$[0,0.1]$  & $0.024254$ & $0.000357$    \\
\hline
$[0.1,0.2]$  & $0.06747$ & $0.001234$    \\
\hline
$[0.2,0.3]$  & $0.103018$ & $0.002539$    \\
\hline
$[0.3,0.4]$  & $0.130306$ & $0.003865$    \\
\hline \hline
\end{tabular}
\end{center}
\end{table}
In the following example, I investigate what happens in the most extreme case, i.e.\ the one achieving the largest error, out of the $20,000$ instances.

\begin{xmpl}
Consider the market values and ratings in Table \ref{rat_simu}

\begin{table}[h!]
\begin{center}
\caption{The ratings and values from the simulation}
\label{rat_simu}
\begin{tabular}{||c|c||c|c|c|c||}
\hline \hline
Item & Market v.\ & $I$ & $II$ & $III$ & $IV$\\
\hline \hline
$A$ &  100 & \ppstab  & \mmstab& \pstab & \ppstab \\
\hline
$B$ & 800 & \mstab  & \mstab & \nstab & \mmstab \\
\hline
$C$ & 460 & \ppstab  & \mstab & \pstab & \nstab\\
\hline
$D$ & 772 & \mstab  & \mstab & \ppstab & \ppstab\\
\hline
$E$ &  432 & \pstab  & \ppstab& \ppstab & \nstab \\
\hline
$F$ & 166 & \pstab  & \nstab & \mstab & \mmstab \\
\hline
$G$ & 631 & \mmstab  & \pstab & \nstab & \mmstab\\
\hline
$H$ & 435 & \mstab  & \mstab & \nstab & \mstab\\
\hline \hline
\end{tabular}
\end{center}
\end{table}
When $K=\sqrt[4]{3/2}$, the following allocation is returned
\begin{equation}
\label{alloc:simul}
z^e=
\kbordermatrix{
     & $A$ & $B$ & $C$
      & $D$ & $E$ & $F$ & $G$ & $H$ 
     \\
     $I$ & 0.340 & 0 & 1 & 0 & 0.355 & 1 & 0 & 0.187 
     \\
     $II$ & 0 & 0 & 0 & 0 & 0.645 & 0  & 1 & 0
      \\
     $III$ & 0 & 1 & 0 & 0 & 0 & 0 & 0 & 0.813
     \\
     $IV$ & 0.660 & 0 & 0 & 1 & 0 & 0 & 0 & 0
}
\end{equation}
If $\sqrt[4]{3/2}$ is the correct value for $K$, agent $III$ receives only items rated three stars -- the highest ratings for those goods. If $K$ grows, but the allocation remains the same, the normalized utility of those goods decreases (because the same agent assigned 5 stars to other goods). Instead all the other agents received items rated 4 or 5 stars -- and their normalized utility  increases with $K$. The utility levels, together with the value of the ``true'' Egalitarian allocation for the corresponding $K$, are shown in Figure \ref{fig:worstcase}. Clearly, allocation \eqref{alloc:simul} is not optimal for $\sqrt[4]{3/2}$ and the sub-optimality grows with $K$.
\begin{figure}[h!]
\begin{center}
\caption{The utility values for allocation \eqref{alloc:simul}, together with the correct Egalitarian value, as a function of $K$.}
\label{fig:worstcase}
\includegraphics[width=0.8\textwidth]{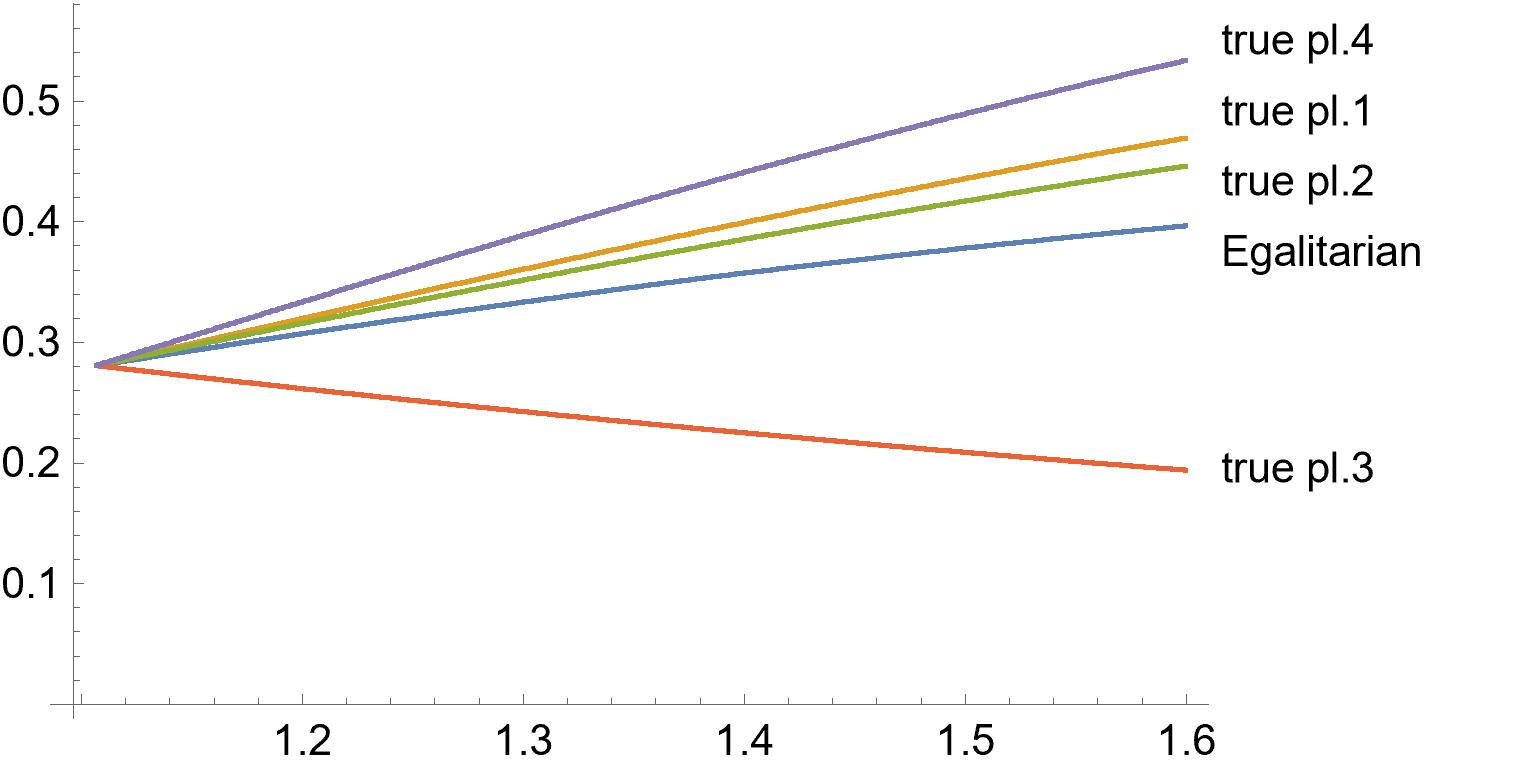}
\end{center}
\end{figure}

\end{xmpl}

When the ``wrong'' magnifying factor is chosen, not only does the allocation become suboptimal, but other properties may fail. For instance, Figure \ref{fig:worstcase} shows that, for a sufficiently large $K$, allocation \eqref{alloc:simul} fails the Fair Share Guarantee property.

A similar simulation study was undertaken to verify the model's robustness against inefficiencies when the parameter $K$ is misspecified. As before, the utility of the optimal allocation with respect to the announced value $K=\sqrt[4]{3/2}$, is computed under a randomly generated true value of the same parameter. As before, 8 items with random market values between $100$ and $1,000$ are divided among 4 agents whose ratings are randomly generated. The utility vector is then compared against a cloud of $20,000$ randomly generated points in the IPS generated by the true $K$. Half of these points correspond to randomly generated integer allocations and their convex combinations, while the remaining half comes from randomly generated fractional allocations in which each $z_{ia}$ is drawn from a uniform $[0,1]$ distributions, and then the allocation is normalized to become feasible. If dominating points are found, the size of domination, averaged over the agents, is 
computed and the maximum of these values is recorded. If no dominating point is found, a null value is recorded. The positive values may underestimate  the domination magnitude and the simulation may report falsely undominated allocations, although the cloud of points is fairly large.

\begin{figure}[h!]
\begin{center}
\caption{Inefficiency from misreported $K$ values:  dominated (blue) vs.\ undominated (orange) solutions}
\label{errordom:plot}
\includegraphics[width=0.8\textwidth]{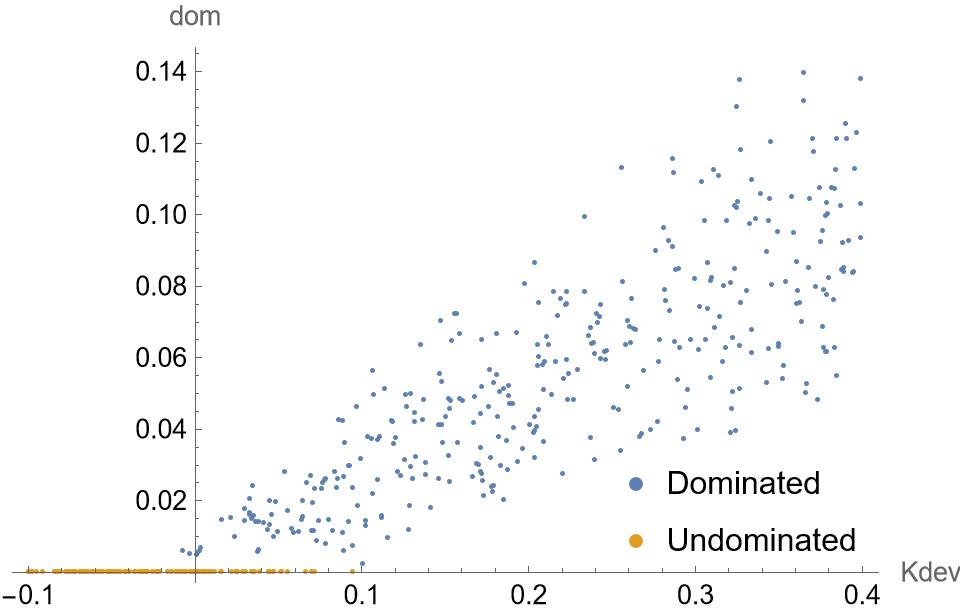}
\end{center}
\end{figure}

Figure \ref{errordom:plot} shows the outcome of 500 simulations, with randomly generated goods' marked values and agents' ratings. Dominated solutions are indicated by blue points: the $x$-coordinate indicates the size and direction of the deviation between the announced and the true value of $K$, while the $y$-coordinate indicates the estimated average domination size. Orange points on the $x$-axis show the undominated cases.

For positive deviations in the value of $K$, the plot shows a predictable behavior in which the size of domination grows linearly with the size of deviation and the undominated cases  become sparses and disappear before the deviation size reaches  0.1. The negative deviations, instead, show a quite surprising pattern: with the exception of two cases in which the absolute deviation size is negligible, all the simulations show that the optimal values obtained from misspecified parameters are undominated. This behavior deserves further investigation. 

The simulations show that an improper choice of the rating scale may affect the optimality of the allocation, but it will only marginally distort the utility values. The choice of the magnifying factor is more critical because the sub-optimality of the proposed allocation is proportional to the gap between the true and the proposed value of $K$. The misspecification of the parameter also affects the efficiency of the proposed solution, Therefore, great care must be employed in its value's calibration.

\subsection{Legal applications: Company law}
The  legal workgroup of the European project described 36 different cases where the fair division of an asset among two or more parties is required by law in two different areas: family law (inheritance, divorce) and company law (liquidations, termination of partnerships).  I now illustrate how the procedure illustrated in Section \ref{subs_procedure} can help solve one of these cases regarding the liquidation of a company in which the three partners are entitled to different shares. It is important to notice that the described case took place before the procedure was set up and the project ended before I could interact further with the legal workgroup. Therefore, I interpreted the agents' preferences based on the case description and, since the descriptions of the agents' preferences were rather succinct, I enriched them with some additional elements. The example, therefore, does not describe the actual behaviour of the agents but rather their possible actions, given the reported elements. Here is the case description provided by the legal workgroup.

\begin{quote}
$I$, $II$ and $III$ concluded a partnership contract in 2006, agreeing to contribute their work and/or property to achieve a common objective – a small carpentry factory and a store for selling goods. They had different stakes/contributions that determined their shares as joint owners. Person $I$ was a carpenter with experience especially in kitchens and bedrooms. He contributed equipment (valued at 35,000 euros) and of course  his “know-how” and experience. Person $II$ had business premises large enough for the factory and for the store, and this was his contribution. Person $III$ contributed 30,000 euros in cash. After the financial crisis, the business began to deteriorate, so person $II$ proposed changing the purpose of their business to stocking and selling electronic appliances that would be directly imported from China. $II$ still thinks that he is the only one who can decide about the purpose of the business premises. $I$ was disappointed because they did not need him or his work anymore. $III$ only cares about profit. The content of their common asset (joint ownership) changed during the decade. They bought new machinery but they also had a special website for selling furniture with the possibility of online interior design as an additional service. To set up this website, they had to spend 4,500 euros and they pay 1,200 euros monthly for software licences and website maintenance fees.
They decided to dissolve the joint ownership and the first step that the court had to make was determining their shares. The court decided that $I$ has 3/9, $II$ has 5/9 and $III$ 1/9 of the business. By determining  their shares, joint-ownership was transformed into co-ownership.
Upon the dissolution of co-ownership (in May 2016), the assets consist of all of the above mentioned items but also includes new machinery (valued at 20,000 euros, store items valued at 30,000 euros, and a profit of 15,000 euros). 

In the process of partitioning co-ownership, $I$ wants all the machinery, but also a part of the property where the factory was located because he wants to continue running the same business by himself. $II$ wants a part of the profits to start with his idea and all the  business premises. He is also interested in the website because he wants to sell online. $III$ is interested in money alone and proposes to sell the business as a whole.
\end{quote}

In addition to the given description, I further assume that the three partners agree on a value of 25,000 euros for the website. Moreover, the agents agree to leave money as one of the disputed items. Table \ref{rat_cl} gives a list of the items and their value and ratings by the partners, compatible with their statements.
\begin{table}[h!]
\begin{center}
\caption{The ratings for the company law case.}
\label{rat_cl}
\begin{tabular}{||c|c||c|c|c||}
\hline \hline
Item & Market v.\ & $I$ & $II$ & $III$ \\
\hline \hline
1. Equipment\ & 35k & \nstab  & \mstab & \mmstab \\
\hline
2. Business Pr.\ & 70k & \ppstab  & \ppstab & \mmstab \\
\hline
3. Machinery & 20k & \ppstab  & \mmstab & \mmstab\\
\hline
4. Store Items& 30k & \mstab  & \mstab & \mstab \\
\hline
5. Website & 25k & \mstab & \ppstab & \mmstab \\
\hline
6. Money & 15k & \mstab  & \pstab & \ppstab \\
\hline
\end{tabular}
\end{center}
\end{table}
By setting $K=\sqrt[4]{3/2}$, the central ratings are
\[
\bar{r}_I=3.6594 \qquad \bar{r}_{II}=3.6374 \qquad    \bar{r}_{III}=1.5270
\]
The procedure returns the following solution, with two split goods, each contended between two agents.
\[
z^e=
\kbordermatrix{
     & 1. \textrm{Eq} & 2. \textrm{BP} & 3. \textrm{Ma}
      & 4. \textrm{SI} & 5. \textrm{We} & 6. \textrm{Ca} 
     \\
     $I$ & 1 & 0.1861 & 1 & 0 & 0 & 0 
     \\
     $II$ & 0 & 0.8139 & 0 & 0.9298 & 1 & 0 
      \\
     $III$ & 0 & 0 & 0 & 0.0702 & 0 & 1
}
\]
The indices in Table \ref{char_cl} characterize the division.

\begin{table}[h!]
\begin{center}
\caption{Characterization of the solution in the company law case.}
\label{char_cl}
 \begin{tabular}{||c|c|| c|c|c||} 
 \hline
 Agent & $w_i$  & $\mu_i(z^e)$ & $\sigma_i(z^e)$ & $\rho_i(z^e)$  \\
 \hline\hline
 $I$ & 1/3   & 68.067 & 1.0472 &   +0.3629
\\ 
 \hline
$II$ & 5/9   & 109.826 & 1.0138 &   +0.6827
\\ 
 \hline
 $III$ & 1/9  & 17.107 & 0.7896 &   +3.1492
\\ 
 \hline
\end{tabular}
\end{center}
\end{table}
The MSE ratio ranks the agents in the order $I$, $II$ and $III$, indicating that agent $I$ gets a monetary share, weighted with the corresponding entitlement, which is larger than that of agent $II$. This, in turn, is larger than that of agent $III$. The unequal treatment is compensated by a reversed order for the RD indices: The allocation provides agent $III$ with a gain of 3.1492 over the central rating. In turn, the gain of agent $II$ with respect to the central rating is only 0.6827, and that of agent $III$ is even smaller.
 Notice that the satisfaction that $III$ gets for receiving the most treasured good is compensated by a lower market value for the received goods.

 \section{The procedure for two agents}
 When the division takes place between two agents only, Brams and Taylor's Adjusted Winner (AW) procedure can be used. The procedure asks the two agents to distribute a fixed number of points -- or utilities --  among the contended items. Items are then ordered figuratively on a line according to the utilities' ratios. Finally, a splitting point is tentatively sought, so that equality in utility is obtained by assigning goods on the sides of the point, one to each agent and by a proper allocation of  the items on the point which may require at most one split item. For more details on the procedure, we refer to Chapter 4 of Brams and Taylor \cite{bt96}.
 
In order to use AW, ratings should be converted into utilities, by using the power rating formula \eqref{ratuti}, and then normalized. The resulting utility values will be expressed as positive real numbers -- not integers. This preliminary work is not needed, except for the real numbers relaxation, because items can be directly arranged on the ordered line according to the rating differences. In fact, when the power rating model holds,
 \[
 \frac{u_{1a}}{u_{2a}}=\frac{m_a K^{r_{1a}-q-1}}{m_a K^{r_{2a}-q-1}}=K^{r_{1a}-r_{2a}}\qquad \mbox{for any } a \in G
 \]
and goods keep the same order on the line -- whether utilities' ratios or rating differences are considered.

As in the general case, the appropriate indices (UM and RD) measure the quality of the optimal allocation in terms of monetary value and rating difference.

\subsection{Legal Application: Divorce}
I examine another instance provided by the legal workgroup of the European project. As in the previous application, the project conclusion prevented me from interacting further with the legal workgroup. I therefore include some additional descriptions regarding the agents'  likes and dislikes, and the ratings reflect my interpretation of the agents' intentions, rather than their recorded behaviour. This is the case as reported by the legal workgroup:

\begin{quote}
$W$, wife of $H$, asks for the statement of cessation of the civil effects of the marriage, three years having passed since the judgment of personal separation.

The goods in common are:
\end{quote}
\begin{enumerate}
\item An apartment, used as a family home, worth $1,500,000$ euros;
\item An apartment in a seaside resort worth $1,250,000$ euros;
\item A prestigious building, inherited by the couple through testamentary disposition, worth $1,750,000$ euros;
\item Valuable furniture (works of art) contained in the aforementioned buildings worth $550,000$ euros;
\item Two cars worth $60,000$ and $50,000$ euros, respectively;
\item A vintage car, worth $170,000$ euros;
\item Company equity investments worth $750,000$ euros;
\item A sum of money equal to $1,500,000 $ euros.
\end{enumerate}
\begin{quote}
The spouses are both professional financial operators in the risk capital market and are involved in several types of entrepreneurial activities. For this reason, both have an interest in retaining company holdings.
The wife is also interested in the valuable furniture  for herself, as part of her entrepreneurial activity involves the buying and selling of works of art.
For his part, $H$ requires the assignment of works of art and vintage cars, as he is a collector.
\end{quote}

To better outline the preferences, I further assume that the wife is interested in the family house and has some interest in the seaside resort apartment, while the husband has agreed to live in the inherited  apartment.

Money (item 8) can either be considered as an item of the division or it can be distributed  in equal parts between the parties. In the previous example, I considered the first option, while here I choose the second alternative to point out the fact that the procedure returns a satisfactory division without money transfers.

\begin{figure}[h!]
\begin{center}
\caption{The items' group arrangement for the divorce example}
\label{div_groups}
\includegraphics[width=0.9\textwidth]{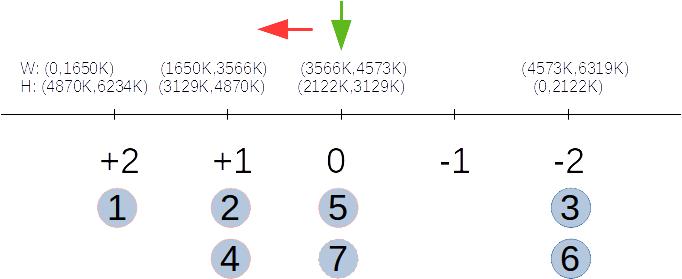}
\end{center}
\end{figure}

\begin{table}[h!]
\begin{center}
\caption{The ratings for the divorce case}
\label{rat_div}
\begin{tabular}{||c|c||c|c||c|c||}
\hline \hline
Item & Market v.\ & $W$'s ratings & $W$'s utility & $H$'s ratings & $H$'s utility \\
\hline \hline
1. Family Apt.\ & 1500k & \pstab & 1650k & \mstab & 1364k \\
\hline
2. Seaside Apt.\ & 1250k & \nstab & 1250k & \mstab & 1136k \\
\hline
3. Inherited Apt.\ & 1750k & \mstab & 1591k & \pstab & 1925k \\
\hline
4. Furniture & 550k & \ppstab & 666k & \pstab & 605k \\
\hline
5. Two Cars & 120k & \mmstab & 99k & \mmstab & 99k \\
\hline
6. Vintage Car & 170k & \mstab & 155k & \pstab & 197k \\
\hline
7. Equity Inv.\ & 750k & \ppstab & 908k & \ppstab & 908k \\
\hline
\end{tabular}
\end{center}
\end{table}
Based on the short description, I set the ratings of the two parties in Table \ref{rat_div}.
Figure \ref{div_groups} shows the assignment of the goods to groups and the utility ranges for the two agents. For each point, two ranges are computed: the utility of goods on the left (right, resp.) for agent 1 (agent 2, resp.), excluding and including the goods located at the point. Starting from a central group (typically, the one in which the difference is zero, if it is non-empty), the procedure verifies whether the two ranges cross. If they do not, the splitting point is moved in the direction of reducing the distance between the two ranges. When the two ranges finally cross, an allocation that assigns all the items to the left (right, resp.)  of the splitting point to agent 1 (agent 2, resp.). The items at the point are allocated so that utilities are equalized. If necessary, one item is split between the agents.

For $K=\sqrt[4]{3/2}$, the central ratings are
\[
\bar{r}_W=3.3892 \qquad \bar{r}_H=3.2320 .
\]

Starting from the items in group 0, I move leftward because the range for agent W is above that of agent H. At ``$+1$'' the two ranges cross, and equality is attained by assigning items 2 and 4 so as to obtain an Egalitarian allocation. Since this can be done in two ways while splitting one item, I get the following optimal allocations:
\begin{gather*}
z_1^e=
\kbordermatrix{
     & 1. \textrm{FA} & 2. \textrm{SA} & 3. \textrm{IA} & 4. \textrm{Fu} & 5. \textrm{2C} & 6. \textrm{VC} & 7. \textrm{Eq}
     \\
     W & 1 & 0.8366 & 0 & 1 & 0 & 0 & 0
     \\
     H & 0 & 0.1634 & 1 & 0 & 1 & 1 & 1
}
\\
z_2^e=
\kbordermatrix{
     & \phantom{1. \textrm{FA}} & \phantom{2. \textrm{SA}} & \phantom{3. \textrm{IA}} & \phantom{4. \textrm{Fu}} & \phantom{5. \textrm{2C}} & \phantom{6. \textrm{VC}} & \phantom{7. \textrm{Eq}}
     \\
     W & 1 & 1 & 0 & 0.6968 & 0 & 0 & 0
     \\
     H & 0 & 0 & 1 & 0.3032 & 1 & 1 & 1
}
\end{gather*}
The indices in Table \ref{char_div}  help define the quality of the division as perceived by the two agents:
\begin{table}[h!]
\begin{center}
\caption{Characterization of the solution for the divorce case.}
\label{char_div}
 \begin{tabular}{||c||c| c||c| c||} 
 \hline
 Agent & $\mu_i(z_1^e)$ & $\rho_i(z_1^e)$ & $\mu_i(z_2^e)$ & $\rho_i(z_2^e)$ \\
 \hline\hline
 W & 3095.73 k &  +  0.4756 & 3133.21 k & +  0.3568
\\ 
 \hline
H & 2994.25 k &  + 0.8043 & 2956.76 k & + 0.9286
\\ 
 \hline
\end{tabular}
\end{center}
\end{table}

In both solutions, the larger market share obtained by $W$ is offset by a slightly lower RD index. I note that the sum of money given to the two agents at the beginning of the procedure can be used to assign the only split item to one of the agents. In both solutions, it seems reasonable for $W$ to buy the smaller share originally assigned to $H$.

 \subsection{What the Agents Should Know}
Whatever the number of agents is, an important issue for implementing the procedure is  the degree of knowledge that the agents should have about the division process before they participate in it. Every detail of the procedure should be public knowledge among the agents, though it is hard to imagine each of the participants being fully aware of every mathematical detail of the algorithm that returns the allocation.

On a more practical level, we can think of a list of clues to be given to the agents in order to make them more aware of the rating process. These indications are not meant as a full replacement of the procedure's description, nor they are meant as axioms. The two-agent setting, with the horizontal line described in step $c)$ of the procedure and pictured in Figure \ref{div_groups} helps explain why these clues work.
\vskip0.3cm
\noindent
{\bf Indications for the agents}
\begin{enumerate}
\item Agents should assign high ratings to the goods they wish most for themselves, and low ratings for the goods they are ready to leave to somebody else.
\item Suppose an agent has tentatively decided on a set of ratings for all the goods. Before passing those ratings to the mediator, the agent may review the ratings with the following indications in mind:
\begin{enumerate}
\item If an agent decides to raise (lower, resp.)\ the rating of a good -- while keeping all the other ratings fixed -- the chances of receiving the good increase (decrease, resp.). This holds because the good changes position on the horizontal line of AW. For instance, if agent 1 increases the rating, the good moves leftward on the horizontal line, and the increase in the rating indicates the movement distance. We use the term ``chance'', because the allocation depends on the difference of the two ratings -- not just on the single agent's rating. The statement can be made more rigorous by assuming that the rating of the other agent is generated by a fixed probability distribution. The probability of getting the good increases with the item's rating.
\item If an agent increases or decreases the ratings of all the goods by the same amount (provided this is allowed by the tentative ratings) nothing changes. This happens because all goods are moved in the same direction and by the same distance on the horizontal line.  
\end{enumerate}
\item At the same time, obtaining goods with higher ratings has a cost: if there are two Egalitarian solutions, then the solution that contains more goods (or parts of them) with higher ratings has a lower market value. 
\end{enumerate}
The last indication is formally justified by the following result, which holds for any number of agents.
\begin{prop}
\label{AvsBprop}
Suppose that for one agent $i \in N$, $r_{iA} > r_{iB}$, $A$ and $B$ being two of the contended goods. Furthermore, suppose that $z'$ and $z''$ are two Egalitarian allocations with 
\begin{equation}
\label{AvsB1}
z'_{iA} \geq z''_{iA}, \: z'_{iB} \leq z''_{iB} 
\end{equation} with at least one of the inequalities strict. Moreover
\begin{equation}
\label{AvsB2}
z'_{ia}=z''_{ia} \quad \mbox{for every }a \neq A,B.
\end{equation}

Then $\mu_i(z') < \mu_i(z'')$.
\end{prop}
\begin{proof}
Since $z'$ and $z''$ are Egalitarian allocations in the same problem, their normalized utility must coincide.  Thus $\bar{U}_i(z')=\bar{U}_i(z'')$ and, by \eqref{AvsB2},
\[
m_A z'_{iA} K^{r_{iA}-\bar{r}_i} + m_B z'_{iB} K^{r_{iB}-\bar{r}_i} = m_A z''_{iA} K^{r_{iA}-\bar{r}_i} + m_B z''_{iB} K^{r_{iB}-\bar{r}_i}.
\]
Dividing both terms by $K^{r_{iB}-\bar{r}_i}$ and rearranging the terms, it becomes
\[
(m_A z'_{iA} - m_A z''_{iA}) K^{r_{iA}-r_{iB}} =m_B z''_{iB} -  m_B z'_{iB}.
\]
By $\eqref{AvsB1}$ both terms are positive and, since $r_{iA} > r_{iB}$ holds,
\[
m_A z'_{iA} + m_B z'_{iB} < m_A z''_{iA}  + m_B z''_{iB} .
\]
This, together with \eqref{AvsB2}, implies $\mu_i(z') < \mu_i(z'')$.
\end{proof}
In the legal application, two solutions, $z^e_1$ and $z^e_2$, are found. On comparing the two solutions, W has a larger part of good 4 and a smaller part of good 2 in $z^e_1$ and good 4 has a higher rating than good 2 for W. All the hypotheses of Proposition \ref{AvsBprop} are satisfied and, as expected, $\mu_W(z^e_1)>\mu_W(z^e_2)$. The same is true for H, but with the roles of $z^e_1$ and $z^e_2$ inverted and, therefore, $\mu_H(z^e_2)>\mu_H(z^e_1)$.

\section{Concluding Remarks and Further Work}

This work illustrates a procedure for allocating divisible goods among agents in such a way that both the agents' preferences and the goods' market values contribute to defining the fairness criterion. The procedure is simple, as agents rate the goods on a simple discrete range, and is computationally fast, since only linear programming techniques are involved, and this keeps the number of split items to the minimum needed to achieve perfect equitability in the agents' utilities. 

The procedure has been used in an applied project which counts people in professional roles from many fields, many of them with little or no mathematical background. As was to be expected, though, the applied work raises many new research questions. Several open problems have been pointed out in the previous section. Other directions include:
\begin{enumerate}[a)]
\item {\em The choice of the parameter $K$.} In the present model, the addition of a star to  a good's rating increases its utility by the factor $K > 1$. By setting $K=\sqrt[4]{3/2}$, I obtained quite reasonable results in the examined applications. Can an optimal value for $K$ be defined based on theoretical and/or behavioural and experimental observations?

\item {\em Increased Model Flexibility.} The DPR utility model is assumed to work with the same parameters for all the agents. Agents may have different discernment abilities. A refinement of the present model should include personalized parameters for each agent.

\item{Analysis of Special Cases} Can the procedure be made easier for some instances? For instance, when all the goods have the same value, does the procedure coincide with AW?

\item {\em Divisibility Constraints.} The model assumes all goods to be divisible. We have already seen that this is not  a major problem even for inherently indivisible goods, because a good's split can be mimicked by means of ownership shares or by monetary compensations among agents. 

In any case, imposing one or more goods to be allocated in their entirety to one of the agents can easily be incorporated as additional integer constraints in \eqref{opt_prob}. The new solution, however, may fail to satisfy perfect equality in the agents' utilities that the described procedure guarantees.

A remedy for this problem is to define new properties that apply to the context of mixed divisible and indivisible allocations of goods. An example of this change of perspective is provided by the recent work of Bei et al.\ \cite{bllll20}, where a novel notion of envy-freeness is found that is guaranteed to exist in the hybrid situation just described.

\item {\em Division of goods and bads.} The list of items to be divided may contain liabilities (``bads'' in the economic jargon) such as debts, obligations or duties to perform. 

The recent works by Bogomolnaia et al. \cite{bmsy17a} and \cite{bmsy17b} reveal that introducing bads is not a simple corollary of the case for goods, but surprising and unexpected results are quite common. It would be interesting to extend the procedure to the mixed case of goods and bads, making use of the already known paradoxes. 

\item {\em Market value and the Nash/Competitive solution}. Finding a relationship between the Nash/Competitive solution and the market value, using the DPR utility model or any other proposal would incorporate the many strengths of the solution with the practicality of tying together the agents' subjective satisfaction with money.
\end{enumerate}

\section*{Disclaimer}
This paper has been produced with the financial support of the Justice Programme of the European Union (Grant Agreement No. 766463, Call: JUST-AG-2016-05, Project Leader: Prof.\ Francesco Romeo). The contents of this paper are the sole responsibility of the author and can in no way be taken to reflect the views of the European Commission.




\end{document}